\documentclass[proof]{pasj00}

\begin{document}

\SetRunningHead{Y. Takeda et al.}{Beryllium Abundances of Red Giants}
\Received{2014/03/27}
\Accepted{2014/06/27}

\title{Spectroscopic Study on the Beryllium \\ 
Abundances of Red Giant Stars
\thanks{Based on data collected at Subaru Telescope,
operated by the National Astronomical Observatory of Japan.}
}

\author{
Yoichi \textsc{Takeda}\altaffilmark{1}
and
Akito \textsc{Tajitsu}\altaffilmark{2}
}
\altaffiltext{1}{National Astronomical Observatory, 2-21-1 Osawa, Mitaka, Tokyo 181-8588}
\email{takeda.yoichi@nao.ac.jp}
\altaffiltext{2}{Subaru Telescope, 650 N. A'ohoku Place, Hilo, HI 96720, U.S.A.}
\email{tajitsu@subaru.naoj.org}



\maketitle

\begin{abstract}
An extensive spectroscopic study was carried out for the beryllium abundances 
of 200 red giants (mostly of late G and early K type), which were determined 
from the near-UV Be~{\sc ii} 3131.066 line based on high-dispersion spectra 
obtained by Subaru/HDS, with an aim of investigating the nature of surface Be 
contents in these evolved giants; e.g., dependence upon stellar parameters, 
degree of peculiarity along with its origin and build-up timing.
We found that Be is considerably deficient (to widely different degree 
from star to star) in the photosphere of these evolved giants by 
$\sim$~1--3 dex (or more) compared to the initial abundance.
While the resulting Be abundances ($A$(Be)) appear to weakly depend
upon $T_{\rm eff}$, $\log g$, [Fe/H], $M$, $age$, and $v_{\rm e}\sin i$, 
this may be attributed to the metallicity dependence of $A$(Be) 
coupled with the mutual correlation between these stellar parameters, 
since such tendencies almost
disappear in the metallicity-scaled Be abundance ([Be/Fe]).
By comparing the Be abundances (as well as their correlations
with Li and C) to the recent theoretical predictions based on sophisticated
stellar evolution calculations, we concluded that such a considerable extent/diversity 
of Be deficit is difficult to explain only by the standard theory of first dredge-up 
in the envelope of red giants, and that some extra mixing process (such as rotational 
or thermohaline mixing) must be responsible, which presumably starts to operate 
already in the main-sequence phase. This view is supported by the fact that appreciable 
Be depletion is seen in less evolved intermediate-mass B--A type stars near to 
the main sequence.
\end{abstract}

%


\section{Introduction}

Beryllium (Be) is an astrophysically important light element like lithium (Li), 
which can be used as a probe of envelope mixing or as a proxy of stellar parameters
such as age or rotational velocity, since it is destroyed when conveyed into
hot deep interior owing to its fragility to nuclear reaction (burned 
in comparatively low temperatures of $\gtsim 3.5 \times 10^{6}$~K).
While it is not easy to determine the abundance of this element, 
since the strong resonance Be~{\sc ii} doublet at the UV region of $\sim 3131$~$\rm\AA$ is 
essentially the only abundance indicator available, a number of spectroscopic studies have 
been accumulated over these several decades (see, e.g., the Proceedings papers 
collected in Charbonnel et al. 2010 and the references therein), thanks to the 
availability of large telescopes built on ideal sites of high atmospheric transparency.

Nevertheless, most Be abundance studies published so far 
have been directed to main-sequence (or turn-off/subgiant) stars of mostly FGK types, 
and little is known about evolved stars such as red giants 
despite that several trials have been reported.
That is, earlier-time studies (e.g., Boesgaard et al. 1977 for Hyades giants;
De Medeiros et al. 1997 for Li-rich giants; Castilho et al. 1999 for Li-rich giants) 
appear to be less reliable in the quantitative sense because of the insufficient 
line list (the blending effect on the Be~{\sc ii} 3131.06 line does not seem to 
be properly taken into account) as well as due to comparatively lower data quality,
though all of them suggested a tendency of Be depletion. Meanwhile, more recent analyses 
by Melo et al. (2005) (for Li-rich giants) and Smiljanic et al. (2010) (several G-type 
giants in the moderately-young open cluster IC~4651) are considered to be more credible 
in this respect. Unfortunately, they failed to determine $A$(Be)\footnote{
We define $A$(E) as the logarithmic abundance for an element E in the usual
definition; i.e., $A$(E) $\equiv \log [N({\rm E})/N({\rm H})] + 12$.} 
of these giants, for which only the upper limit values could be estimated, 
since the Be~{\sc ii} 3030--3131 doublet lines were found to be very weak
almost near or below the detection limit (i.e, being essentially overwhelmed 
by blending of neighboring lines).
Their work suggests that determining Be abundances in red giants
(which seem considerably depleted in the general sense) is a difficult and 
challenging task, to say the least.

Takeda, Sato, and Murata (2008; hereinafter referred to as Paper I)
carried out an extensive spectroscopic study on 322 late G and early K giants
(targets of Okayama Planet Search Program) with various luminosities
corresponding to mass range of $\sim$~1.5--5~$M_{\odot}$, and determined 
the stellar parameters as well as the chemical abundances of 17 elements 
in a consistent manner. Given that the properties of
these red giants are well established, they make a good sample for investigating 
the connection between the surface abundance change of fragile light elements 
and the evolution-induced mixing in the envelope. 

Along with this line, Liu et al. (2014) very recently investigated the behavior 
of Li abundances for a large sample of 378 G/K giants (where all 322 targets in 
Paper I are included), and found the following characteristics:\\
--- Li is heavily depleted in the surface of these stars. Actually,  Li line is 
invisible in $\sim 1/3$ of the sample stars, for which only the upper limits 
of $A$(Li) could be assigned (ranging from $\ltsim 0$ to $\ltsim 1$).\\ 
--- Even for the detection cases, $A$(Li) ranges between $\sim 0.5$ and $\sim 2$, 
which means by $\sim$~1.5--3~dex underabundant relative to the solar-system 
abundance of $A_{\rm s.s.}$(Li) = 3.31 (Anders \& Grevesse 1989).\\ 
--- Li line is undetected in all 23 planet-host stars, which implies
that Li depletion tends to be enhanced by the existence of planets.\\
--- Since the extent of deficiency is too large to be explained by the conventional 
stellar evolution theory for the first dredge-up in the red-giant phase 
(by as much as $\sim 1.5$~dex), a considerable portion of Li depletion 
may have taken place already at the main-sequence phase.  

Thus, it is interesting to examine for these red-giant stars whether 
a similar trend is observed for Be, from the viewpoint of similarity in the
characteristics of these two elements (though Li is burned in somewhat lower 
temperature at $T \gtsim 2.5 \times 10^{6}$~K). 

Admittedly, as mentioned above, it is not easy to extract information on Be 
abundances from the very weak and blended Be~{\sc ii} feature in considerably 
Be-depleted cases (expected for evolved giants), especially when 
theoretical and observed spectra are compared simply by eye-judgement.   
Conveniently, however, we have experiences of Be abundance determination for 
a large sample of solar analog stars (Takeda et al. 2011, hereinafter referred 
to as Paper II), where an automatic solution-search algorithm to accomplish 
the best spectrum fitting in the neighborhood of the Be~{\sc ii} 3131.066 line 
was applied to establish the Be abundance. Equipped with this technique,
we decided to challenge studying the behaviors of Be abundances 
for a large number (200) of late G and early K-type giants, based on 
high-dispersion spectra obtained by observations with Subaru Telescope.
The purpose of this paper is to report the results of this investigation.

The remainder of this article is organized as follows. 
After describing our observations and data reduction in section 2, 
we explain the procedures of our analysis (spectrum-fitting, equivalent-width 
derivation, and upper-limit estimation) in section 3.  
Section 4 is devoted to discussing the characteristics of the resulting
Be abundances, especially in connection with stellar parameters and
Li abundances, and the conclusion is summarized in section 5.
Besides, we describe in appendix 1 additional numerical experiments,
which were conducted on artificial spectra to study the nature of 
errors in our Be abundance determination.
In appendix 2 are also presented the results of our supplementary 
Be~{\sc ii} 3130--3131 doublet analysis for 5 late B through late A-type stars, 
which was carried out to get information on the surface Be abundances of 
intermediate-mass stars before reaching the red-giant phase. 

\section{Observational Data}

All of the 200 targets in this investigation (see table 1) were selected from 
322 late G or early K giants, which were already studied in detail in Paper I. 
The observations  were carried out on 2013 July 17 and 19 (UT) with 
the High Dispersion Spectrograph (HDS; Noguchi et al. 2002) placed at 
the Nasmyth platform of the 8.2-m Subaru Telescope, 
by which high-dispersion spectra covering $\sim$~3000--4600~$\rm\AA$ could 
be obtained with two CCDs of 2K$\times$4K pixels in the standard Ub setting 
with the blue cross disperser. 
We used the slit width of $0.''6$ (300 $\mu$m) and a binning of 2$\times$2 pixels,
which resulted in a spectrum resolving power of $R \simeq 60000$.  
The typical integrated exposure times was $\sim$~5--10~min for each star 
(typically $V \sim 5$). 

The reduction of the spectra (bias subtraction, flat-fielding, 
scattered-light subtraction, spectrum extraction, wavelength calibration,
co-adding of frames to improve S/N, continuum normalization) was 
performed by using the ``echelle'' package of 
the software IRAF\footnote{IRAF is distributed
    by the National Optical Astronomy Observatories,
    which is operated by the Association of Universities for Research
    in Astronomy, Inc. under cooperative agreement with
    the National Science Foundation.} 
in a standard manner. 
Typical S/N ratios of $\sim$~50--100 (estimated from counts of
photo-electrons; cf. table 1) were attained at the position of 
Be~{\sc ii} 3130--3131 doublet lines in the finally resulting spectra.

\section{Determination of Be Abundances}

\subsection{Spectrum Fitting Analysis}

The strategy and the procedure of our Be abundance determination are 
almost the same as in Paper II, where we focus on the Be~{\sc ii} 3131.066 line 
(weaker one of the doublet). Though this line is blended with Fe~{\sc i} 
3131.043, it is still superior to the stronger Be~{\sc ii} 3130.421 line
which is severely contaminated by the strong V~{\sc ii}+OH line feature
at $\sim 3130.3$~$\rm\AA$. This situation is illustrated in figure 1,
where observed spectra in the neighborhood of Be~{\sc ii} doublet for two
representative stars are simply compared with the theoretical spectra 
computed with various Be abundances.

Thus, our analysis is based on the synthetic spectrum fitting applied to 
the narrow region of 0.7~$\rm\AA$ width centered at 3131~$\rm\AA$ 
Practically, we used the stellar spectrum analysis tool ``MPFIT'', 
which was developed based on Kurucz's (1993) ATLAS9/WIDTH9 program and 
has a function of establishing the spectrum-related parameters 
(elemental abundances, macrobroadening parameters, radial velocity, 
etc.) by automatically searching for the best-fit solutions 
without any necessity of precisely placing the continuum level 
in advance (Takeda 1995).\footnote{
In the present application to near-UV region heavily crowded with many spectral lines
(and thus finding the line-free window is hopeless), it is essential to introduce
not only the normalization constant ($C$) but also the tilt-adjustment parameter 
($\alpha$) in comparing the observed flux ($f_{\lambda}$) with the theoretical flux 
($F_{\lambda}$) in order to accomplish a satisfactory fit. 
That is,  the right-hand side of equation (1) in Takeda (1995) is redefined as 
$\sum^{}_{} \{\log f_{\lambda} - \log F_{\lambda} - C - \alpha 
(\lambda - \lambda_{\rm min})/(\lambda_{\rm max}- \lambda_{\rm min})\} / N$,
and the best values of $C$ and $\alpha$ are iteratively established, 
where $\lambda_{\rm max}$ and  $\lambda_{\rm min}$ are the maximum and minimum wavelength
of the relevant region. 
} 

We interpolated Kurucz's (1993) grid of ATLAS9 model atmospheres in terms 
of $T_{\rm eff}$, $\log g$, and [Fe/H] (which were taken from Paper I) 
to generate the atmospheric model for each star. 
Then, given the photospheric model along with the microturbulence ($v_{\rm t}$,
also taken from Paper I), we determined for each star the abundances 
of three elements [$A$(Be), $A$(Ti), and $A$(Fe)],\footnote{The contributions 
of the lines of other elements (E) than these three were formally 
included by assuming the solar abundances scaled with the metallicity; 
i.e., $A$(E) = $A_{\odot}$(E) + [Fe/H]. Note also that we did not vary
$A$(Nb) which was fixed (unlike the case in Paper II), 
since the Ti~{\sc ii} 3130.810 line is much stronger
than the Nb~{\sc ii} 3130.780 line (the former dominating the Ti+Nb feature) 
in the spectra of red giants.} along with the macrobroadening parameter 
($v_{\rm M}$; $e$-folding half-width of the Gaussian macrobroadening
function, $f_{\rm M}(v) \propto \exp [(-v/v_{\rm M})^{2}]$) and the
radial velocity shift, by applying the MPFIT program 
to the observed spectrum in the 3130.65--3131.35~$\rm\AA$ region
with the same line data as given in table 1 of Paper~II.
As in Paper II, we assumed LTE (local thermodynamical equilibrium) throughout 
this study.\footnote{As done in Paper II (cf. the Appendix 
therein), we examined how the non-LTE effect is important on Be abundance determination 
for models with different $T_{\rm eff}$ (4500, 5000, 5500~K) and $\log g$  (1.5 and 3.0).
We then found that the non-LTE correction ($\Delta \equiv A_{\rm NLTE}-A_{\rm LTE}$)
is always positive (i.e., the non-LTE effect acts as a line-weakening)
and quite sensitive to $\log g$ (but not to $T_{\rm eff})$; i.e., 
$\Delta \ltsim$~0.1--0.2~dex ($\log g = 1.5$) and $\Delta \ltsim$~0.05 ($\log g = 3.0$).
Accordingly, the extent of the non-LTE effect is somewhat larger for the present case 
of low-gravity red giants, as compared to the case of solar analogs in Paper II. 
Nevertheless, since corrections of such amount (on the order of $\sim$~0.1~dex
at most) are not significant compared to the large diversity of $A$(Be) 
($\gtsim 2$~dex), we may neglect them without any serious problem.}

The solution of $A$(Be) (along with those of other parameters) turned out 
to converge for 130 stars (out of 200 targets in total).
When $A$(Be) could not be determined (i.e., the solution did not converge 
because the Be~{\sc ii} line is too weak), we neglected its contribution by 
assuming $A$(Be) = $-9.99$ and repeated the iteration to accomplish the fit.
For these undetermined cases (70 stars), we estimated the upper limit
values of $A$(Be) as described in the next subsection. 
The theoretical spectrum corresponding to the finally established parameter solutions 
is compared with the observed spectrum for each star in figure 2, where we can see 
that the agreement is mostly satisfactory.

It should be remarked that this is actually a very difficult analysis, since 
the contribution of the Be~{\sc ii} line is not apparent but barely detectable 
only as a subtle second-order effect (even for the cases of well-established solutions). 
This situation is demonstrated in figure 3, where the observed spectra in the 
relevant region for stars with similar parameters but with appreciably different 
$A$(Be) solutions are compared with each other. As seen from this figure, the extent 
of Be abundance reflects on (i) the slight wavelength shift (toward redder for 
higher $A$(Be)) of the Fe~{\sc i} 3131.043 + Be~{\sc ii} 3131.066 line feature 
and (ii) the strength ratio of this Fe~{\sc i}+ Be~{\sc ii} feature to the 
neighboring Fe~{\sc i} 3131.25 line (larger for higher $A$(Be)). 
However, the differences are too small to be confidently discernible
by human eyes, though they can be discriminated by such a computer-based numerical 
judgement as adopted by us. Accordingly, the resulting solutions are generally 
very delicate and vulnerable to slight spectrum defect/noise, which means that 
they tend to suffer rather large uncertainties (especially near to the detection limit)
though such ambiguities are difficult to quantify.

\subsection{Equivalent Width and Upper Limit}

As in Paper II, we prefer using equivalent widths (rather than the resulting 
abundance solutions themselves) which are easy to handle and useful 
in many respects.

We evaluated the (imaginary) equivalent width of the Be~{\sc ii} 3131.066 line 
corresponding to the detection limit as 
$ew_{\rm Be II\; 3131}^{\rm DL} \simeq k \times$~FWHM/(S/N),
where $k$ is a factor we assumed to be 2 according to our 
experience, S/N is the signal-to-noise ratio of the relevant spectrum 
($\sim$~50--100), and FWHM was estimated from $v_{\rm M}$ as 
FWHM~$\simeq 2\sqrt{\ln 2} \; (\lambda v_{\rm M}/c)$ ($c$: velocity of light). 
Typical values of $ew_{\rm Be II\; 3131}^{\rm DL}$ are $\sim$~2--5~m$\rm\AA$ (cf. table 1). 
Regarding 70 stars for which $A$(Be) could not be determined, we estimated its 
upper limit from $ew_{\rm Be II\; 3131}^{\rm DL}$.

Besides, we computed also for the $A$(Be)-established cases (130 stars) 
$EW_{\rm Be II\; 3131}$ and $EW_{\rm Fe I\; 3131}$ ``inversely'' from the solutions 
of $A$(Be) and $A$(Fe) (resulting from the spectrum synthesis analysis) along 
with the adopted atmospheric model/parameters (cf. subsection 3.3 in Paper II). 
Then, considering that the reliability of $A$(Be) may be assessed by 
the relative comparison between $EW_{\rm Be II\; 3131}$ 
and $ew_{\rm Be II\; 3131}^{\rm DL}$, we divided the results into four reliability 
classes (a, b, c, and x):
\begin{itemize}
\item class (a) $\cdots$ reliable solution ($EW_{\rm Be II\; 3131} > 3 \times ew_{\rm Be II\; 3131}^{\rm DL}$) [97 stars]. 
\item class (b) $\cdots$ less reliable solution 
($ 3 \times ew_{\rm Be II\; 3131}^{\rm DL} > EW_{\rm Be II\; 3131} >  ew_{\rm Be II\; 3131}^{\rm DL}$) [28 stars]. 
\item class (c) $\cdots$ unreliable solution ($ ew_{\rm Be II\; 3131}^{\rm DL} > EW_{\rm Be II\; 3131}$) [5 stars].
\item class (x) $\cdots$ undetermined cases (only upper limit) [70 stars]. 
\end{itemize}
The final results of $EW_{\rm Be II\; 3131}$ and $A$(Be) (or upper limits) along with 
the corresponding reliability classes are summarized in table 1. 
It is worth noting that class~(a) results occupy the range of $A$(Be)~$\gtsim -1$
(see, e.g., figure 5), which is consistent with our numerical experiment
(described in appendix 1), implying that Be abundance errors tend to become
significant at $A$(Be)~$\ltsim -1$.

Figure 4a shows a comparison between $A$(Fe)$_{\rm 3131\,fit}$ (derived 
from this fitting) and $A$(Fe)$_{EW}$ (already established in Paper I 
by using a number of Fe lines), which reveals that both correlate 
reasonably with each other, though considerable discrepancies are seen
for several cases. The trend that $A$(Fe)$_{\rm 3131\,fit}$ tends to be
somewhat larger than $A$(Fe)$_{EW}$ toward the metal-rich regime
(where the Fe~{\sc i} 3131.043 line becomes stronger and more saturated)
might be due to the depth-dependent microturbulence increasing with
hight in the low-density atmosphere of red giants (e.g., Takeda 1992),
since we used $v_{\rm t}$ determined from lines of yellow region
(Paper I) while the forming depth of near-UV lines is comparatively higher.

According to figure 4b, where the strengths of Be~{\sc ii} and Fe~{\sc i}
lines at $\sim 3131$~$\rm\AA$ ($EW_{\rm Be II\; 3131}$ and $EW_{\rm Fe I\; 3131}$)
are compared with each other,  we can recognize that $EW_{\rm Be II\; 3131}$ 
widely vary from $\sim 0$ to $\sim 60$~m$\rm\AA$ while the Fe~{\sc i} line 
strength typically distributes around $EW_{\rm Fe I\; 3131} \sim$~60--70~m$\rm\AA$.
As a result, the inequality relation $EW_{\rm Be II\; 3131} \ltsim EW_{\rm Fe I\; 3131}$ 
mostly holds and the Fe~{\sc i} line tends to be predominant over the 
Be~{\sc ii} line in the Fe~{\sc i} + Be~{\sc ii} feature. 

We also examined by using $EW_{\rm Be II\; 3131}$ how the $A$(Be) results 
are sensitive to ambiguities in the adopted atmospheric parameters
($T_{\rm eff}$, $\log g$, and $v_{\rm t}$).
Assuming $\pm 100$~K, $\pm 0.2$~dex, and $\pm 0.2$~km~s$^{-1}$ as typical uncertainties 
in $T_{\rm eff}$, $\log g$, and $v_{\rm t}$ (see subsection 3.1 in Paper I, especially 
the comparison with literature values shown in figures 5--7 therein), we 
found that these perturbations caused changes in $A$(Be) by $\sim \pm 0.04$~dex, 
$\sim \mp 0.12$~dex, and $\ltsim 0.01$~dex (i.e., negligible), respectively.
While the $\log g$-sensitivity is comparatively large (due to the characteristic 
behavior in the line-strength of ionized species), we may generally state 
that errors in the atmospheric parameters are practically insignificant 
for the results of Be abundances.\footnote{
For example, while our $\log g$ values tend to be systematically lower by 
$\sim$~0.2--0.3 dex in comparison with those published by other investigators 
(cf. Fig. 5--8 in Paper I), this influences $A$(Be) only by $\ltsim$~0.2~dex, 
which is not important as compared with the observed large scatter.
}

\section{Discussion}

\subsection{Trend of Be Abundances}

The final results of $A$(Be) derived from our analysis for 200 red giants are plotted 
against $T_{\rm eff}$, $\log g$, $v_{\rm e}\sin i$, [Fe/H], $M$, and $age$ in figures 5a--5f.
In addition, in order to examine the extent of abundance peculiarity (or depletion degree)
while removing the metallicity dependence in the initial abundance, similar plots with 
respect to [Be/Fe] are also shown in figures 6a--6f, where [Be/Fe] is the logarithmic 
Be abundance relative to the solar system (meteoritic) value of 
$A_{\rm s.s.}$(Be) = 1.42 (Anders \& Grevesse 1989) scaled with Fe 
([Be/Fe]~$\equiv$~$A$(Be)$ -1.42 - $[Fe/H]).\footnote{
It may be worth noting that
the initial Be abundance of these intermediate-mass stars of population~I
($-0.7 \ltsim$~[Fe/H]~$\ltsim +0.3$) may not have been simply 
proportional to the metallicity. That is, while the scaling relation between $A$(Be) 
and [Fe/H] (as a result of Galactic chemical evolution) roughly holds 
over a wide metallicity range down to very metal-poor regime,
the slope of $\delta A$(Be)/$\delta$[Fe/H] appears to be smaller than unity 
(i.e., $\sim 0.5$) as far as the metallicity range of disk stars ($-1 \ltsim$~[Fe/H])
is concerned, as discussed in subsection 4.1 of Paper II. Nevertheless,
[Be/Fe] in the present definition is practically sufficient for the present purpose, 
since the metallicity span of our program stars is not large (i.e., only $\ltsim 1$~dex).
It is reasonable to regard the meteoritic result of $A_{\rm s.s.}$(Be) = 1.42 as 
the initial Be abundance of a solar-metallicity star, since figure 8b of Paper II 
suggests $A$(Be)~$\sim 1.5$ for this value. For reference, the solar photospheric
Be abundance was estimated to be $A_{\odot}$(Be) = 1.22 in Paper II, where they
concluded that a moderate decrease of $A$(Be) by $\sim$~0.2--0.3~dex has actually
undergone in the solar envelope, ruling out the possibility of erroneous
underestimation due to (controversial) UV missing opacity. 
}
A close inspection of figure 5 suggests some rough trends regarding the dependence 
on stellar parameters: $A$(Be) tends to slightly increase with $T_{\rm eff}$ (figure 5a), 
$\log g$ (figure 5b), $v_{\rm e}\sin i$ (figure 5c), [Fe/H] (figure 5d), 
and $M$ (figure 5e), while it shows a decreasing tendency with $age$ (figure 5f).

We must recall here, however, that these parameters are not independent but
correlated with each other. According to figure 3 of Paper I, lower mass ($M$) giants  
tend to be of lower luminosity, lower $T_{\rm eff}$, older $age$, and lower [Fe/H], 
while the $M$-dependence of $\log g$ is somewhat complicated (positive correlation 
in the global sense, but anti-correlation for $\sim$~2--3~$M_{\odot}$ stars
belonging to the majority; cf. figure 3e therein). Besides, since the rotational velocity 
markedly slows down with a decrease in $T_{\rm eff}$, $v_{\rm e}\sin i$ systematically
decreases with a decrease in $M$ (cf. figures 10e and 10f in Paper I).
This means that a dependence on one specific parameter may produces a spurious dependence 
on other parameters when they are correlated. In the present case, we consider that these 
apparent trends are essentially attributed to the metallicity dependence 
of $A$(Be) (figure 5d), since these tendencies almost disappear when we use 
[Be/Fe] instead of $A$(Be), as we can recognize in figure 6. 

\subsection{Extent of Be Depletion and Its Implication}

We can see from figure 5 and figure 6 that $A$(Be) shows a considerably 
large dispersion amounting to $\gtsim 2$~dex, and that Be have suffered significant 
depletion (by $\sim$~1--3~dex with widely different degrees from star to star) 
in comparison to the initial Be abundance at the time of star formation.
It would be interesting to compare this observed trend with theoretical predictions.
In figure 6 are also shown the runs of $\log [X(^{9}{\rm Be})/X_{0}(^{9}{\rm Be})]$ 
(logarithmic depletion factor of surface $^{9}$Be atoms relative to the initial composition) 
for 1.5, 2.5, and 4.0~$M_{\odot}$ solar-metallicity ($Z = 0.014$) stars 
in the red-giant phase\footnote{
In order to avoid 
complexity caused by inclusion of near-main-sequence data, we restricted the theoretical 
plots (lines) only to those of the well-evolved red-giant stage satisfying the conditions
of  $T_{\rm eff} < 5700$~K and $age > 10^{7.5}$~yr in figure 6 and figure 8.
See figure 7 for the expected runs of $\log [X(^{9}{\rm Be})/X_{0}(^{9}{\rm Be})]$ during 
the whole evolutionary history. We also checked the results of calculations for 
a somewhat lower metallicity ($Z = 0.004$), but the differences were found 
to be insignificant.  
} 
simulated by Lagarde et al. (2012) based on two different 
treatments of envelope mixing; i.e., standard treatment and treatment 
including rotational and thermohaline mixing. By comparing the observed [Be/Fe] with 
such computed $\log [X(^{9}{\rm Be})/X_{0}(^{9}{\rm Be})]$ in figure 6, we can read 
the following characteristics and implications concerning the envelope-mixing 
process producing surface Be depletion:\\
--- The extent of Be underabundance predicted by the conventional theory is only 
$\ltsim 1.5$~dex and thus quantitatively insufficient to account for the observed 
amount of depletion ($\sim$~1--3~dex), while the inclusion of rotational and thermohaline 
mixing produces more enhanced depletion (typically $\sim $~2--2.5~dex) which is 
closer to the observed tendency and thus comparatively preferable.\\
--- Accordingly, we may state that only the canonical theory of first dredge-up is not 
sufficient, and extra mixing processes have to be additionally taken into consideration 
to explain the surface Be abundance trends of red giants.\\
--- Since such a special mechanism (specifically, rotational mixing matters in this case, 
since thermohaline mixing is restricted to evolved red giants) begins to operate 
already at the main-sequence phase according to Lagarde et al.'s (2012) simulation 
(cf. figure 7), this means that Be anomaly (depletion) must have built up in the 
early-time of stellar evolution before reaching the red-giant phase.\\
--- This consequence is ensured by another related observational fact.
According to our supplementary analysis for five early-type stars of late B though 
late A-type stars (cf. appendix 2), which are progenitors of late G and early K giants 
having masses of $\sim$~2.5--5~$M_{\odot}$, most of them (4 out of 5) indicate significant 
Be deficiencies by $\gtsim$~0.5--1~dex (cf. table 2). This evidence may lend support for 
the scenario that depletion of surface Be begins already in the main-sequence phase. 

\subsection{Comparison with Li and C}

Finally, it is worthwhile to compare the Be abundances resulting from this study with 
those of Li (Liu et al. 2014) and C (Paper I) and to check the consistency with 
theoretical predictions, since both also suffer abundance changes by mixing of 
nuclear-processed products. 

Regarding lithium,  we can confirm that Li and Be share similar characteristics
in several respects, which indicate that these two elements may have experienced
similar depletion history in the envelope of red giants:\\
--- The abundances of Be (derived in this study; based only on class-a results ) 
and Li (from Liu et al. 2014; only ``reliably determined'' results were used) 
tend to show a reasonable correlation with each other, as shown in figure 8a
($A$(Be) vs. $A$(Li)) and figure 8a$'$ ([Be/Fe] vs. [Li/Fe]\footnote{
We define [Li/Fe] $\equiv$ $A$(Li)$ - 3.31 - $[Fe/H], where $A_{\rm s.s.}$(Li) = 3.31
is the solar-system meteoritic abundance of Li (Anders \& Grevesse 1989).
Note that a remark similar to the case of Be (cf. footnote 7) should apply 
also to this case regarding this normalization in terms of the metallicity:
i.e.,, it is not clear whether initial $A$(Li) scales with [Fe/H].
Nevertheless, it is reasonable to assume such a metallicity dependence 
(at $-1 \ltsim$~[Fe/H]) according to chemical evolution calculations (see, e.g., 
figure 9 in Takeda \& Kawanomoto 2005), though its observational confirmation
is difficult because this element tends to be depleted with different degrees
from star to star already in the main-sequence phase. 
}).
Actually, according to our linear-regression analysis applied to these data 
(filled circles in figure 8a and figure 8a$'$, excluding two outliers 
of HD~160781 and HD~212430), we found 
$A$(Be) = $0.81 (\pm 0.16)$ $A$(Li) $- 1.00 (\pm 0.17)$
with a correlation coefficient of $r = 0.66$, and
[Be/Fe] = $0.56 (\pm 0.15)$  [Li/Fe] $- 0.28 (\pm 0.32)$
with a correlation coefficient of $r = 0.54$.
Such a relation between the surface abundances of these two elements is also expected 
from Lagarde et al.'s (2012) theoretical simulation as shown in figure 8a$'$.\\  
--- The fact that Li line was not detected and only upper limit 
of $A$(Li) could be estimated for all the planet-host stars included in 
Liu et al.'s (2014) sample, is similarly seen for the present case of Be. 
That is, $A$(Be) could not be determined for 13 out of 15 planet-harboring 
stars in our sample (cf. table 1), which means that the non-detection 
probability is markedly higher than the case of non-planet-host stars.\\ 
--- Given the existence of such Li--Be correlation, it is understandable that Liu et al. (2014)
derived essentially the same consequence for Li (as we did for Be) regarding the build-up 
timing of abundance anomaly (beginning already at the main-sequence phase).
 
As to carbon, although a correlation between $A$(Be) and $A$(C) might be intuitively
expected since mixing-induced deficit of Be should be accompanied by a decrease of C (due to 
mixing of CN-cycled products), this would not be easy to detect according to 
Lagarde et al.'s (2012) calculation, since [C/Fe] does not monotonically 
correlate with [Be/Fe] (i.e., changing in a somewhat complex manner) and the variation range
of the former is appreciably smaller than the latter as illustrated in figure 8b$'$. 
Thus, it is no wonder that any meaningful correlation is not observed between Be ad C
in figure 8b and figure 8b$'$ (the correlation coefficient is $r \simeq 0.1$, 
based on the class-a results). Still, given that the extents of observed [Be/Fe] and [C/Fe] 
are favorably compared with the theoretical predictions (cf. figure~8b$'$), 
we may state that our results are well reasonable. 

\section{Conclusion}

An extensive spectroscopic study was conducted for establishing the 
Be abundances of 200 red giants (mostly of late G and early K type) 
in order to investigate the behaviors of the surface abundances of 
this fragile element and their implications; e.g., whether they are 
normal or peculiar, how they depend on various stellar parameters, 
how they suffer the depletion process such as evolution-induced 
envelope mixing.

Based on high-dispersion spectra obtained with Subaru/HDS, we analyzed 
the narrow wavelength region in near UV comprising the 
Be~{\sc ii} 3131.066 line (blended with the Fe~{\sc i} 3131.043 line) 
by using the spectrum-synthesis technique along with the automatic 
solution-search approach (already adopted in Paper II), where the 
stellar parameters of target stars were taken from Paper I.

It turned out that this Be~{\sc ii} line is considerably 
weakened (due to the general tendency of Be depletion) and dominated 
by the neighboring Fe~{\sc i} line in most cases, which makes the 
analysis difficult. Actually, while we could somehow arrive at
converged solution of $A$(Be) for 130 stars, its determination failed 
for the rest of 70 stars, for which only upper limits were estimated.

The resulting $A$(Be) was found to slightly depend upon stellar parameters,
in the sense that it tends to decline (i.e., Be deletion being more 
enhanced) somewhat with increasing $age$ as well as with a decrease 
in $T_{\rm eff}$, $\log g$, $v_{\rm e}\sin i$, 
[Fe/H], and $M$. However, since these apparent trends almost disappear 
when use use [Be/Fe] instead of $A$(Be), they are considered to be 
due to the metallicity dependence of $A$(Be) coupled with the mutual dependence
of these stellar parameters. 

We found that $A$(Be) as well as [Be/Fe] show a considerably large dispersion 
amounting to $\gtsim 2$~dex, and that Be has suffered significant depletion 
by $\sim$~1--3~dex (or even more for non-detection cases) compared to
the expected initial abundance.

Consulting Lagarde et al.'s (2012) theoretical simulations, we found that 
the expected extent of Be underabundance based on the conventional theory 
(only by $\ltsim 1.5$~dex) is quantitatively insufficient, while the 
inclusion of rotational and thermohaline mixing predicts more enhanced 
depletion (typically by $\sim $~2--2.5~dex) being closer to the required amount.
This implies that only the canonical thery of first dredge-up is not 
sufficient and extra mixing processes have to be additionally taken 
into consideration. 

Since such special mixing mechanisms begin to operate already at 
the main-sequence phase, appreciable Be anomaly must have built up 
in the early-time of stellar evolution before reaching 
the red-giant phase. This view is also supported by our supplementary 
analysis for late B and A-type stars, for which we found significant 
Be deficiencies by $\gtsim$~0.5--1~dex.

We confirmed a reasonable correlation between the abundances of Be and Li,
which is consistent with the theoretical prediction.  
Besides, since $A$(Be) could not be determined for most (13 out of 15) 
planet-harboring stars, Be depletion appears to be enhanced by the existence 
of giant planets, which was also reported by Liu et al. (2014) for 
the case of Li. Accordingly, given that Li and Be share similar 
characteristics, they may have experienced similar depletion history.

As to carbon, although we could not detect any meaningful correlation between
Be and C (contrary to our naive expectation), this is understandable because 
both correlate (not monotonically but) only in a somewhat complex manner 
with each other. At any event, the extents of observed [Be/Fe] and [C/Fe] 
are reasonably consistent with the theoretical predictions.  

\bigskip

We express our heartful thanks to an anonymous referee for a number of 
valuable comments and suggestions, which were of great help
in improving the contents of this paper. 

\appendix
\section{Accuracy-Check Experiment on Artificial Spectra}

In order to understand the nature of errors involved in our Be abundance 
determination procedure using the synthetic spectrum-fitting
technique, we carried out supplementary numerical experiments.
Adopting $T_{\rm eff} = 4900$~K, $\log g = 2.5$, [Fe/H] = 0.0, 
$v_{\rm t} = 1.5$~km~s$^{-1}$, and $v_{\rm M}$ = 5~km~s$^{-1}$  
as the representative set pf parameters, we first computed theoretical 
spectra for six different beryllium abundances ($A$(Be) = $-1.5$, $-1.0$, 
$-0.5$, 0.0, +0.5, and +1.0), for which the corresponding 
$EW_{\rm Be II\; 3131}$ values are 2.4, 7.3, 20.1, 45.3, 76.1, 
and 104.1~m$\rm\AA$, respectively (while $EW_{\rm Fe I\; 3131}$ 
is 66.2~m$\rm\AA$ in this case). Then, randomly-generated noises of 
normal distribution (coorresponding to S/N = 100, 50, and 20) were 
added to them, and this process was repeated 100 times for each of 
the 18 ($=6\times3$) cases. In this way, 1800 artificial spectra with 
different combinations of ($A$(Be), S/N) were generated. 
Next, we tried Be abundance determination in exactly the same manner
as described in subsection 3.1 for each spectrum. 
From the resulting many $A$(Be) solutions for each ($A$(Be), S/N) combination, 
the average ($\langle A$(Be)$\rangle$) as well as the standard 
deviation ($\sigma$) were derived as presented in table 2, where the number 
of succefullly determinations among 100 trials ($N$) is also given.

Inspecting this table 2, we can see several characteristic trends 
regarding abundance errors in terms of $A$(Be) and S/N. \\
--- The abundance error ($\sigma$) progressively increases with a decrease
in $A$(Be) as well as in S/N (though the averaged abundances are 
almost consistent with the given ones), which is naturally understandable.\\
--- Especially, an appreciable error-enhancement is noticeable when 
$A$(Be) drops below $\sim -1$.\\
---  Even cases where Be abundance is undeterminable (i.e., the iterative 
solution never converged) gradually appear at $A$(Be)$\ltsim -1$.

Accordingly, we may roughly state that $A$(Be)$\sim -1$ is nearly 
the critical border with respect to the reliability of Be abundances
we determined in section 3, in the sense that significant errors 
may be involved in the results of $A$(Be)$\ltsim -1$  while those of 
$A$(Be)$\gtsim -1$ have comparatively higher reliability. This conclusion is
reasonably consistent with the classification defined in subsection 3.2,
because the lower envelope of class~(a) is located around $A$(Be)$\sim -1$
where class~(b) abundances also overlap (cf. figure 5).

\section{Beryllium in Early-Type Stars}

In order to understand the nature of Be depletion observed in G--K giants
of intermediate mass ($\sim$~2--5$~M_{\odot}$), it is important to get 
information on the surface Be abundances of unevolved late B--A stars 
near to the main sequence (corresponding to the relevant mass range).
Although several observational studies on Be abundances of upper-main 
sequence stars were published in 1970--1980s, which were mainly motivated 
by interest on the chemical peculiarities in HgMn stars (where Be tends 
to be conspicuously overabundant in many cases), they were based 
on low-quality spectra of photographic plates or IUE satellite. Actually, 
Be~{\sc ii} 3130--3131 doublet lines could not be detected for normal 
(i.e., non-HgMn) late B--A type stars, for which only upper limits were 
derived, in these pioneering investigations (see, e.g., Boesgaard et al. 1982).

Given this situation, we tried Be abundance determinations for 
five late B--A stars (see table 2 for the list) corresponding to
$\sim$~2--5$~M_{\odot}$ (cf. their positions on the HR diagram shown 
in figure 9), since their high-quality Subaru/HDS spectra were available to us. 
Most of the observational data were obtained on 2013 July 19 (UT),
except for Vega which was observed on 2010 May 25 (UT), with the same 
setting as adopted for our 200 main targets.
Regarding the atmospheric parameters, $T_{\rm eff}$ and $\log g$ were 
determined from the colors of Str\"{o}mgren's $uvby\beta$ 
photometric system with the help of the {\tt uvbybetanew}\footnote{
$\langle$http://www.astro.le.ac.uk/\~{}rn38/uvbybeta.html$\rangle$.}
program (Napiwotzki et al. 1993) as done in Takeda et al. (2009), 
while we assumed reasonable values of $v_{\rm t}$ 
(1, 2, and 4~km~s$^{-1}$)by consulting the empirical 
$v_{\rm t}$ vs. $T_{\rm eff}$ formula (cf. equation (1) 
in Takeda et al. 2009)

Our analysis was done in a similar manner as described in subsection 3.1,
except that spectrum fitting was done in a wider wavelength range
(3128--3132.5~$\rm\AA$) and OH molecular lines were neglected.
We could somehow get the solution of $A$(Be) converged for all 5 stars
as presented in table 2. However, except for 15~Vul exhibiting sufficiently
strong Be~{\sc ii} 3130/3131 doublet, detections of these lines 
are delicate and uncertain for other 4 stars, as shown in figure 10.
As a matter of fact, $EW_{\rm Be II\; 3131}$ (equivalent width of 
the Be~{\sc ii} 3131.066 line; cf. subsection 3.2) corresponding to
the solution of $A$(Be) was evaluated as 1.1, 3.2, 0.6, 76.8, and 4.3~m$\rm\AA$
for $\pi$~Cet, $\alpha$~Lyr, $o$~Peg, 15~Vul, and HD~186377, respectively.
Since the detection limit ($ew_{\rm Be II\; 3131}^{\rm DL}$) is on the order
of several to $\sim 10$~m$\rm\AA$ in this case, the $A$(Be) results for 
these 4 stars ($\pi$~Cet, $\alpha$~Lyr, $o$~Peg, and HD~186377) are subject 
to considerable uncertainties (class (c) according to the classification 
in subsection 3.2) and had better be regarded rather as upper limits.

Keeping this in mind, we conclude the following characteristics
regarding the Be abundances of these stars.\\
--- Given the results that $A$(Be)$\ltsim$~0.8--0.9 ($\pi$~Cet and $\alpha$~Lyr)
 $A$(Be)$\ltsim 0$ ($o$~Peg), and  $A$(Be)$\ltsim -0.4$ (HD~186377),
4 (out of 5) stars of our sample show significant Be deficiencies by 
$\gtsim$~0.5--2~dex to different degrees from star to star, as compared 
to the initial Be abundance of $A$(Be)~$\sim 1.5$ (cf. subsection 4.1).\\
--- Only one star (15~Vul), for which we obtained $A$(Be)$\simeq 1.5$, 
seems to retain the initial Be abundance without any anomaly, 
which is interesting. Yet, we had better bear in mind a possibility 
that this star might have been a HgMn star with overabundant Be
when it was on the main sequence.\\
--- To sum up, while the surface Be abundances of these late B--A stars 
are considerably diversified, we may generally state that Be tends to 
be significantly underabundant, which implies that some kind of 
depletion process takes place already in the main-sequence phase.



\newpage

\setcounter{figure}{0}
\begin{figure}
  \begin{center}
    \FigureFile(80mm,140mm){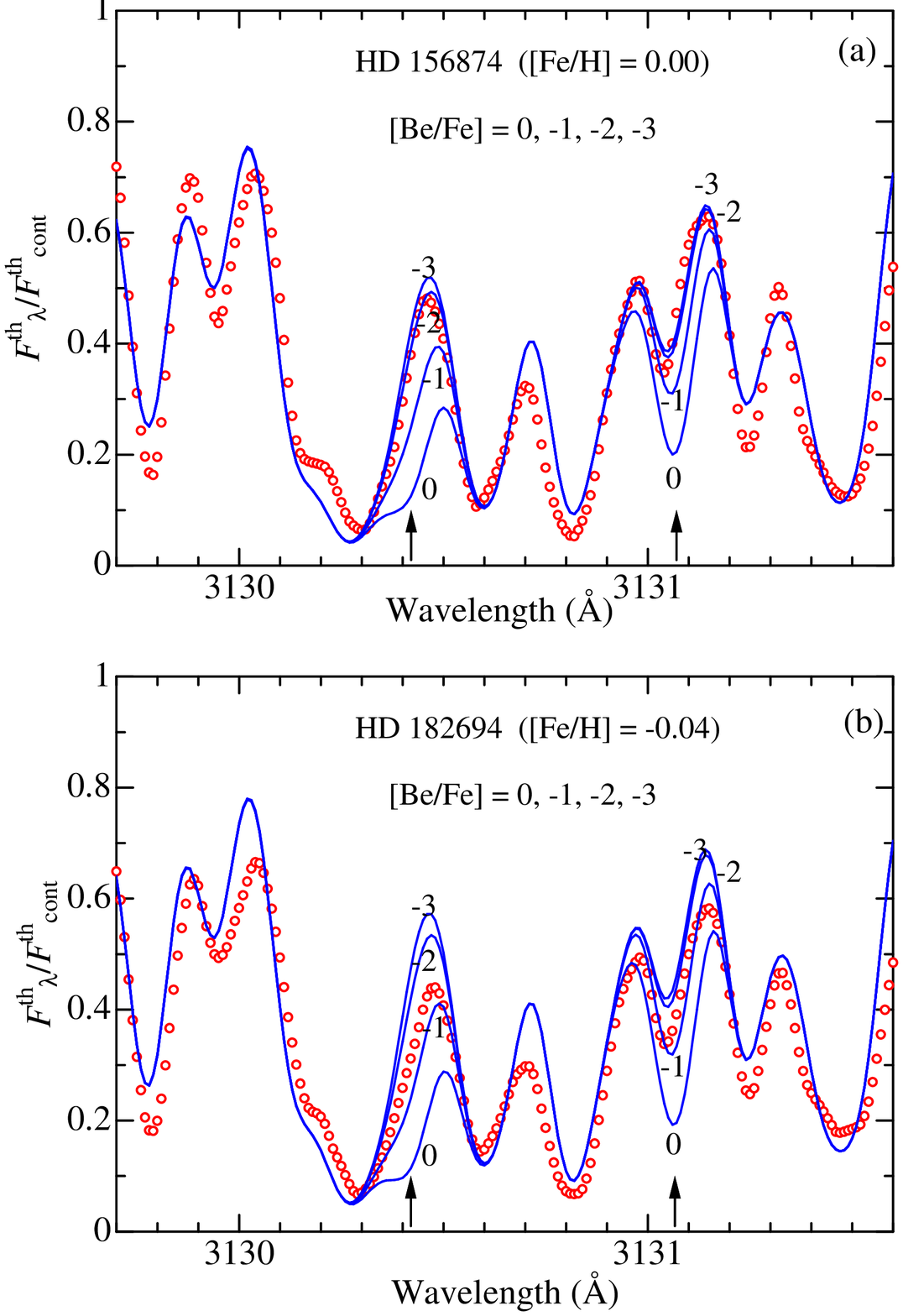}
  \end{center}
\caption{Examples of comparison between the observed spectrum (red open circles; 
normalized by an appropriately assigned continuum level) and the theoretical 
simulation (blue solid lines; $F^{\rm th}_{\lambda}/F^{\rm th}_{\rm cont}$)
in the wavelength region comprising Be~{\sc ii} 3130.421 and 3131.066 lines. 
The upper panel (a) shows the undetermined case (HD~156874; $A$(Be) did not converge) 
and the lower panel (b) corresponds to the determinable case (HD~182694; $A$(Be) converged).  
The theoretical spectrum is synthesized basically with the metallicity-scaled solar 
abundances ([X/Fe]~=~0) and broadened with $v_{\rm M}$ = 5~km~s$^{-1}$, but four 
different $A$(Be) values ([Be/Fe] = 0, $-1$, $-2$, $-3$) are assigned to show 
its effect.
}
\end{figure}

\setcounter{figure}{1}
\begin{figure}
  \begin{center}
    \FigureFile(160mm,200mm){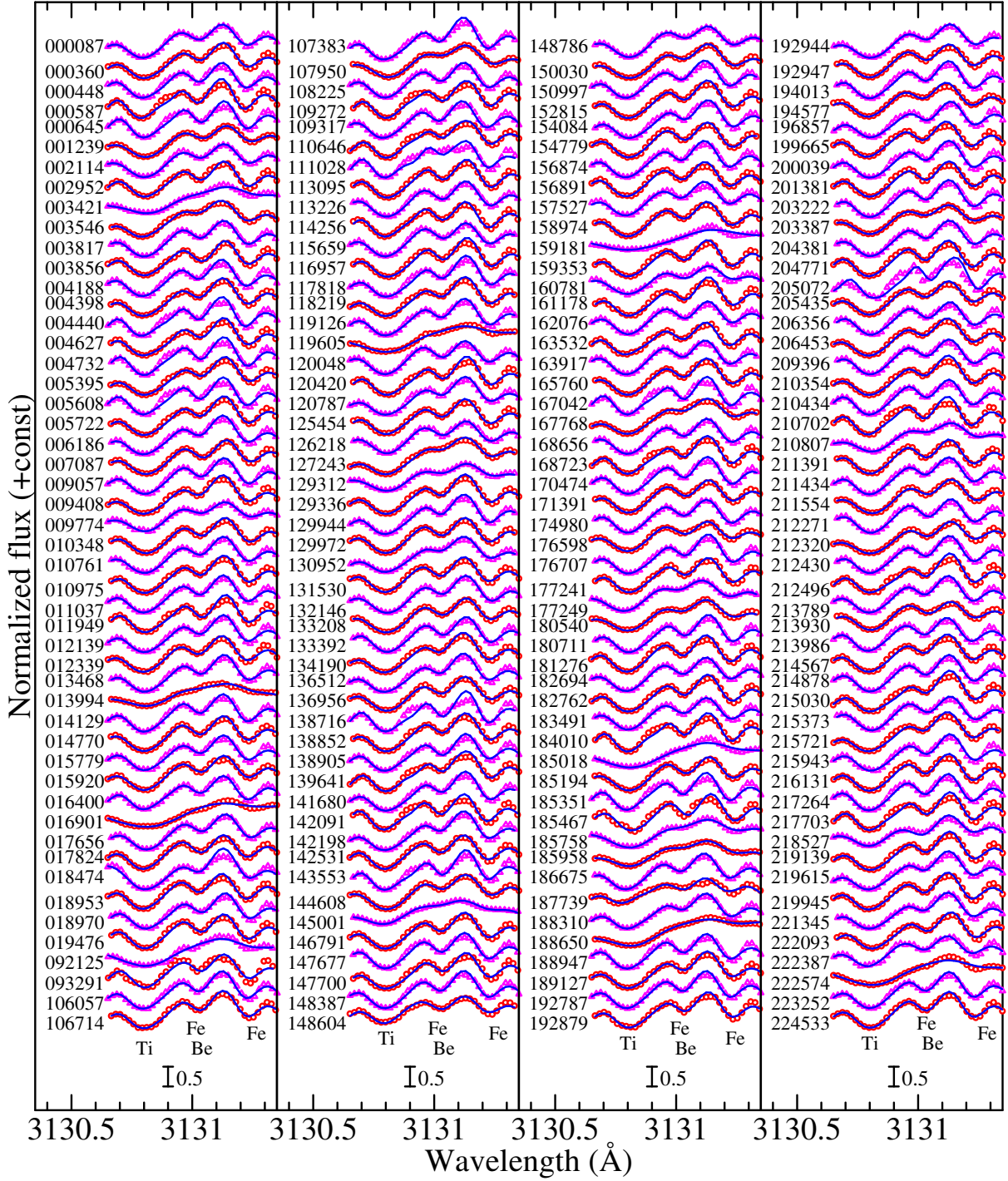}
  \end{center}
\caption{
Synthetic spectrum fitting in the 3130.65--3131.35~$\rm\AA$ region 
for all of the 200 targets. The best-fit theoretical spectra
are shown by solid lines, while the observed data are plotted
by open circles. A vertical offset of 0.5 is applied to each relative to
the adjacent ones. Each of the spectra are arranged in the
increasing order of HD number (indicated on the left to each 
spectrum). The wavelength scale of each spectrum is adjusted to 
the laboratory system.
}
\end{figure}

\setcounter{figure}{2}
\begin{figure}
  \begin{center}
    \FigureFile(80mm,140mm){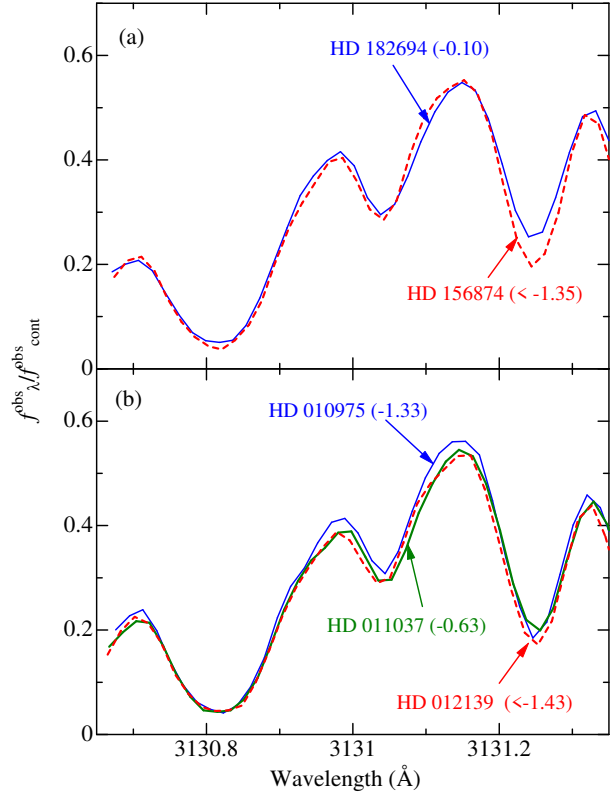}
  \end{center}
\caption{
Demonstration that only very subtle changes are recognizable in the observed 
spectra (in the Be~{\sc ii} 3131.07 region) for stars with similar parameters but 
with appreciably different $A$(Be). (a) HD~182694 and HD~156874 (see also 
figure 1). (b) HD~10975, HD~11037, and HD~12139. The corresponding $A$(Be) 
values are also indicated after the star names. The continuum of the observed
spectra ($f^{\rm obs}_{\rm cont}$) used here for the normalization was
evaluated in reference to the theoretical continuum ($F^{\rm th}_{\rm cont}$),
while using the solutions of $C$ and $\alpha$ (cf. footnote 3) derived as
by-products of spectrum fitting.
}
\end{figure}

\setcounter{figure}{3}
\begin{figure}
  \begin{center}
    \FigureFile(80mm,120mm){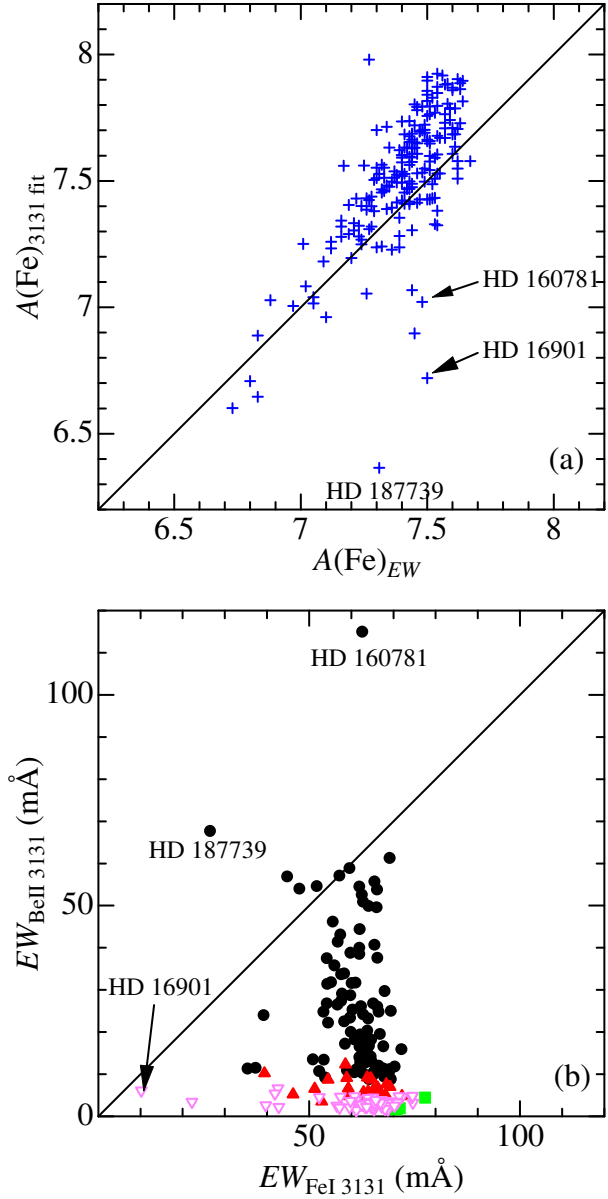}
  \end{center}
\caption{
(a) Correlation of the Fe abundance derived from the synthetic 
spectrum fitting analysis in the 3130.65--3131.35~$\rm\AA$ region (this study)
with that determined based on $EW$s of many Fe lines (cf. Paper I). 
(b) Comparison of $EW_{\rm Fe I \;3131.043}$ and $EW_{\rm Be II\; 3131.066}$ 
(equivalent widths inversely computed from fitting-based 
abundance solutions of Fe and Be). 
Symbols indicate the relevant reliability class  for the $A$(Be) solution 
(cf. caption in figure 5). Note that upper-limit values of $EW_{\rm Be II\; 3131}$ 
are plotted here for undetermined cases (class x; pink downward triangle)
}
\end{figure}

\setcounter{figure}{4}
\begin{figure}
  \begin{center}
    \FigureFile(130mm,170mm){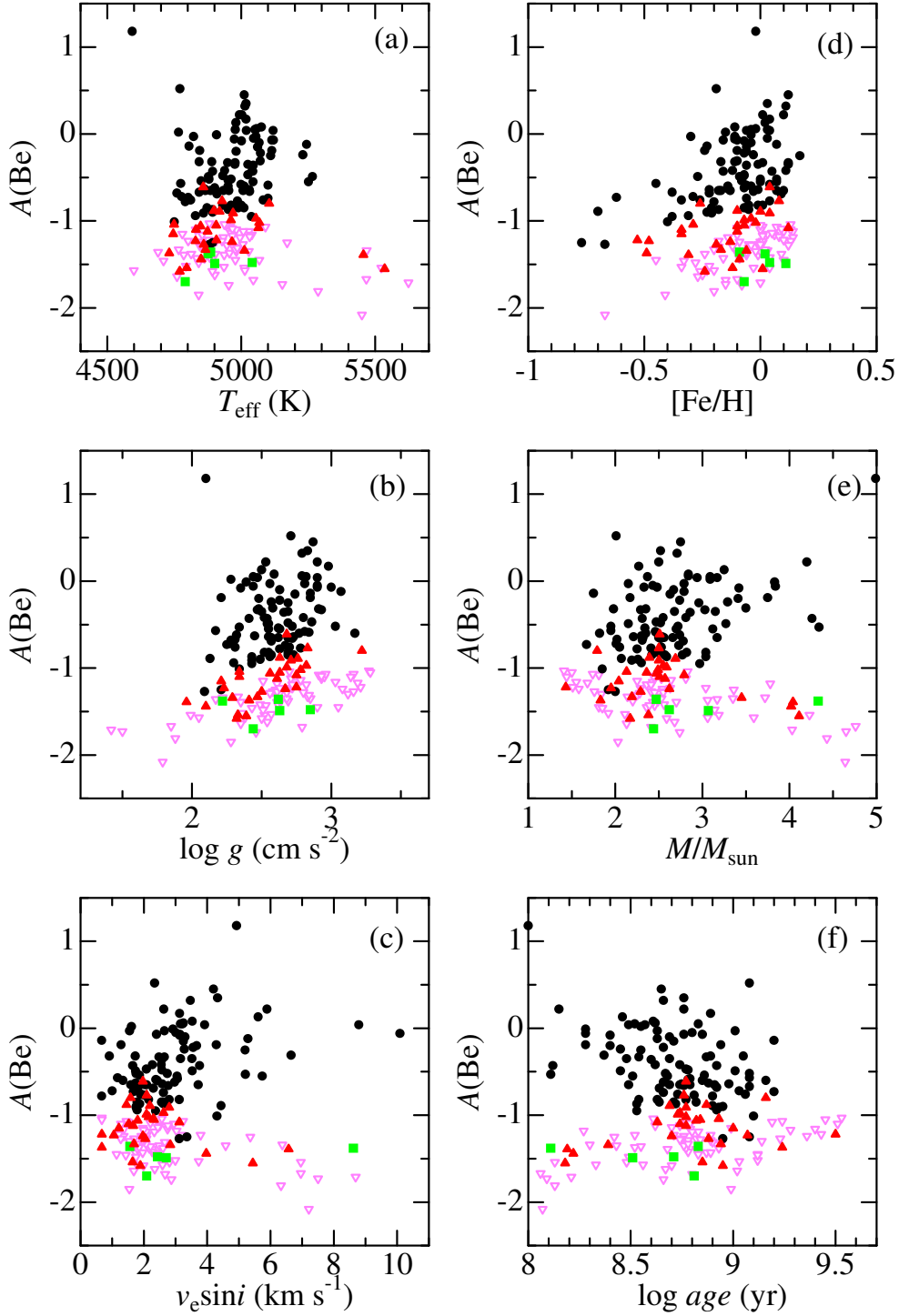}
  \end{center}
\caption{
Beryllium abundances of the program stars plotted against six 
stellar parameters presented in table 1:
(a) $T_{\rm eff}$, (b) $\log g$, (c) $v_{\rm e}\sin i$, (d) [Fe/H],
(e) $M$, and  (f) $age$.
Black filled circles $\cdots$ reliable abundances (class a);
red filled triangles $\cdots$ less reliable abundances (class b)
green squares $\cdots$ unreliable results near/below the detection limit (class c);
pink downward triangles $\cdots$ upper-limit values for the undetermined 
cases (class x). See subsection 3.2 for more details regarding this classification.
}
\end{figure}

\setcounter{figure}{5}
\begin{figure}
  \begin{center}
    \FigureFile(130mm,170mm){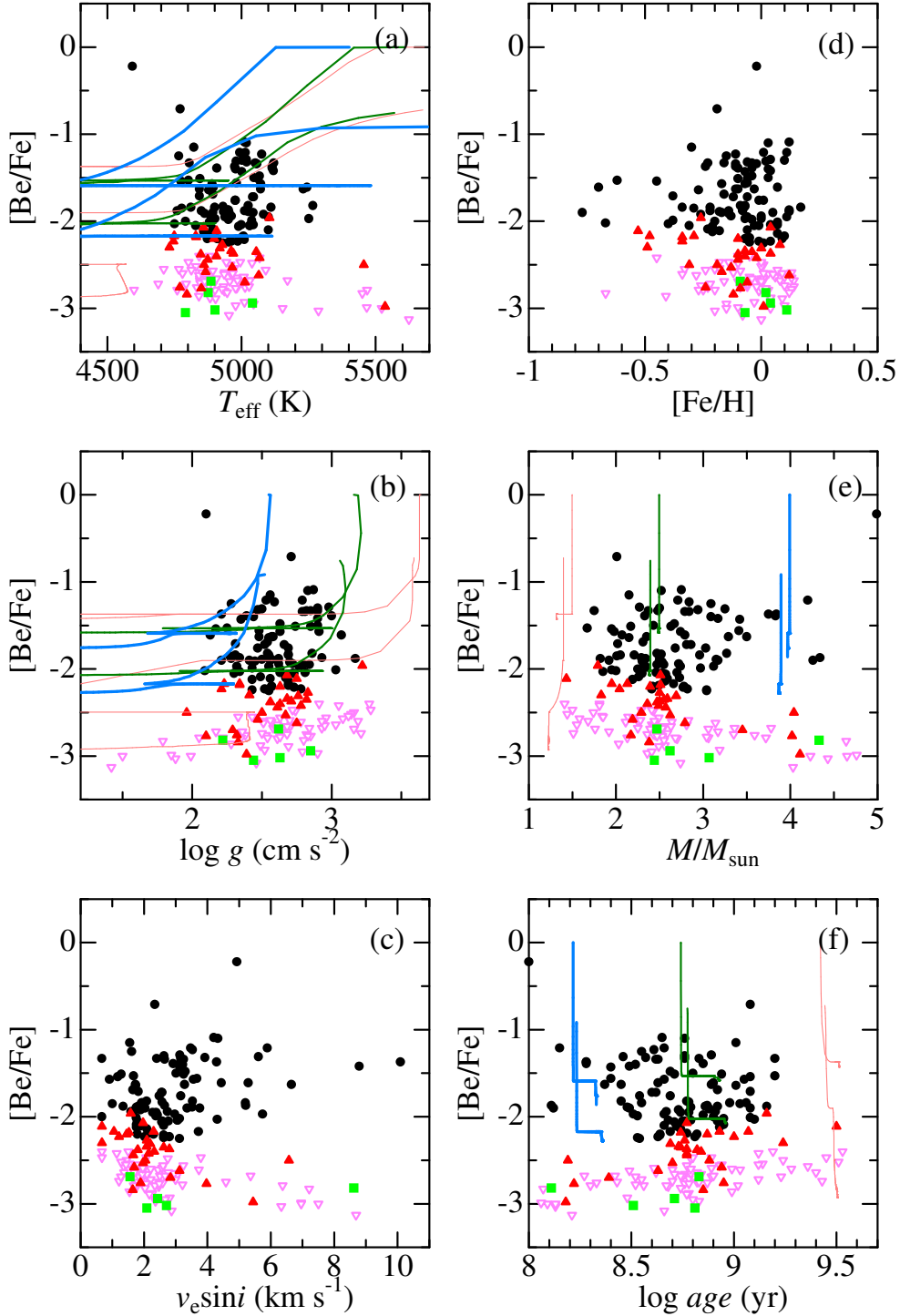}
  \end{center}
\caption{
[Be/Fe] (metallicity-scaled logarithmic Be abundances 
relative to the solar-system composition) of the program stars 
plotted against six stellar parameters, arranged in the same manner 
as in figure 5. See the caption of figure 5 for the meanings of the symbols.
In panels (a), (b), (e), and (f), Lagarde et al.'s (2012) theoretically 
simulated results of $\log [X(^{9}{\rm Be})/X_{0}(^{9}{\rm Be})]$ 
(logarithmic surface abundance of $^{9}$Be relative to the initial composition), 
in the red-giant phase for the solar metallicity ($Z = 0.014$) are also depicted for comparison, 
where stellar masses are discriminated by line thickness (thin orange line, normal 
green line, and thick blue lines correspond to 1.5, 2.5, and 4.0~$M_{\odot}$, respectively).
Note that two kinds of curves are shown corresponding to different treatments 
of envelope mixing; i.e., standard treatment and treatment including rotational 
and thermohaline mixing. Although these two sets are drawn in the same line-type,
they are discernible as the latter generally shows appreciably lower Be abundances 
by 0.5--1.0~dex as compared to the former. In panel (e), the curves of the latter set
are slightly shifted by $-0.1 M_{\odot}$ in order to avoid degeneracy.
}
\end{figure}

\setcounter{figure}{6}
\begin{figure}
  \begin{center}
    \FigureFile(100mm,100mm){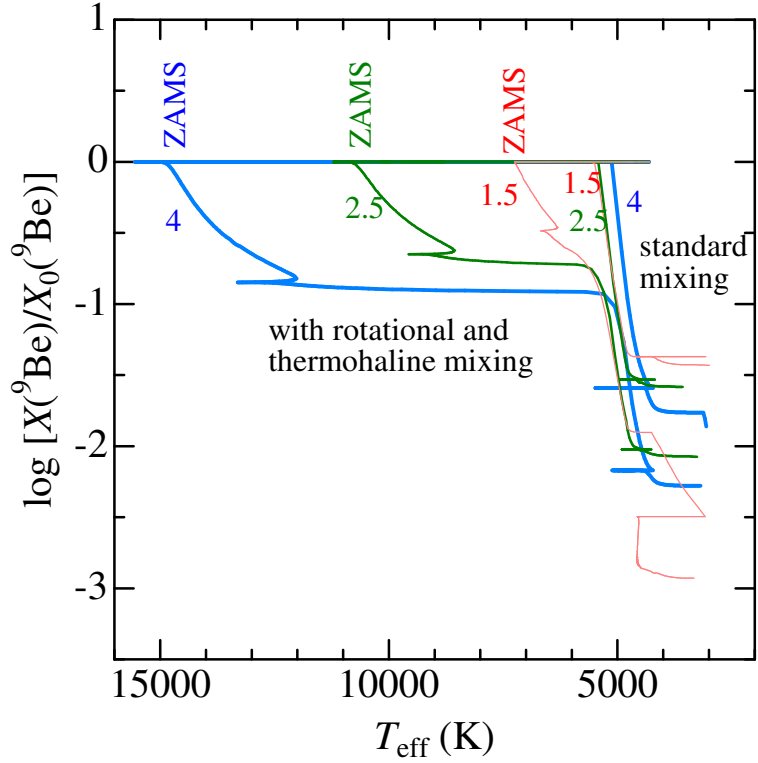}
  \end{center}
\caption{
Run of $\log [X(^{9}{\rm Be})/X_{0}(^{9}{\rm Be})]$
(logarithmic surface abundance of $^{9}$Be
relative to the initial composition), theoretically simulated along 
the whole evolutionary sequence (Lagarde et al. 2012),
plotted against $T_{\rm eff}$.
As in figures 6 (and figure 8), shown here are the results corresponding to three 
stellar masses of 1.5, 2.5, and 4.0~$M_{\odot}$ for two different treatments 
of envelope mixing: (i) standard treatment, (ii) treatment including rotational 
and thermohaline mixing. See the caption of figure 6 for the meaning of the 
line-type.
}
\end{figure}

\setcounter{figure}{7}
\begin{figure}
  \begin{center}
    \FigureFile(130mm,130mm){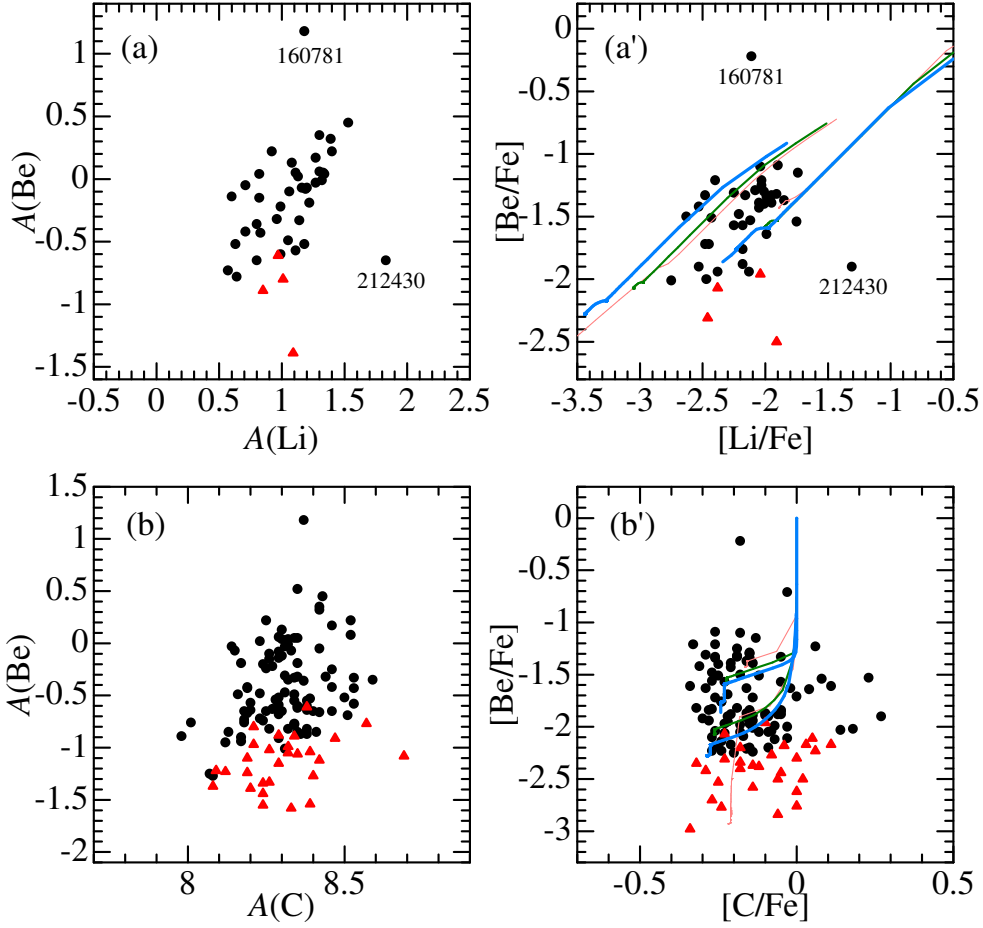}
  \end{center}
\caption{
Beryllium abundances plotted against the abundances of other two 
key elements; Li (upper panels) and C (lower panels).
The left panels give the direct comparison of $A$(X) (X is Li or Be 
or C), while the right panels show the mutual correlations of [X/Fe] 
(metallicity-scaled logarithmic abundance relative to the solar 
composition [for C] or solar-system composition [for Li and Be]).
The data of $A$(Li) are taken from Liu et al. (2014; only the results 
which they classified as ``reliably determined'' were used),
while those of $A$(C) are from Paper I.  
Note that we show only the reliable (class a) 
as well as less reliable (class b) $A$(Be) values here 
(disregarding unreliable results as well as upper-limit cases). 
The meanings of the symbols are the same as in figure 5.
In the right panels (a$'$ and b$'$), the correlations of 
Lagarde et al.'s (2012)  theoretically simulated results of 
$\log [X(^{7}{\rm Li})/X_{0}(^{7}{\rm Li})]$, 
$\log [X(^{9}{\rm Be})/X_{0}(^{9}{\rm Be})]$, and
$\log [X(^{12}{\rm C})/X_{0}(^{12}{\rm C})]$
are also shown (see the caption in figure 6 for the meaning
of the line-type).
}
\end{figure}

\setcounter{figure}{8}
\begin{figure}
  \begin{center}
    \FigureFile(100mm,100mm){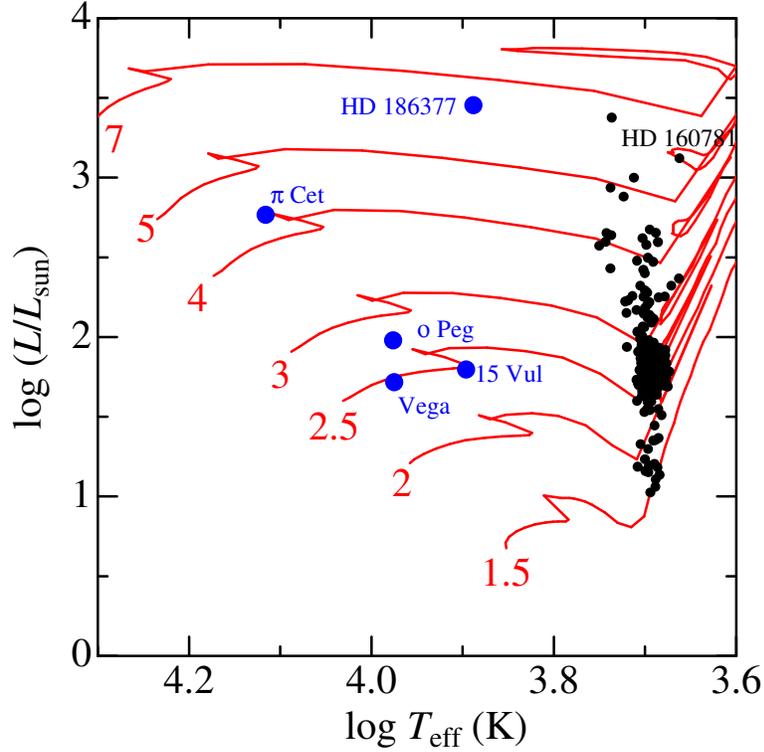}
  \end{center}
\caption{
Five late~B--late~A stars (the targets of our supplementary analysis) are plotted 
on the theoretical HR diagram ($\log (L/L_{\odot})$ vs. $\log T_{\rm eff}$) 
by blue filled circles, where 200 program stars of red giants  
are also shown by black dots for comparison.
The effective temperature ($T_{\rm eff}$) was determined from $uvby\beta$ 
photometry as described in the text, and the bolometric luminosity ($L$) was 
evaluated from the apparent visual magnitude, Hipparcos parallax 
(van Leeuwen 2007), Arenou, Grenon, and G\'{o}mez's (1992) interstellar 
extinction map, and Flower's (1996) bolometric correction. 
Theoretical evolutionary tracks corresponding to the solar metallicity 
computed by Lejeune and Schaerer (2001) for seven different initial masses 
(1.5, 2, 2.5, 3, 4, 5, and 7~$M_{\odot}$) are also depicted by red lines.
Along with the five targets, the position of HD~160781 (K giant with
an exceptionally high Be abundance of $A = +1.18$) is also indicated.
}
\end{figure}

\setcounter{figure}{9}
\begin{figure}
  \begin{center}
    \FigureFile(130mm,170mm){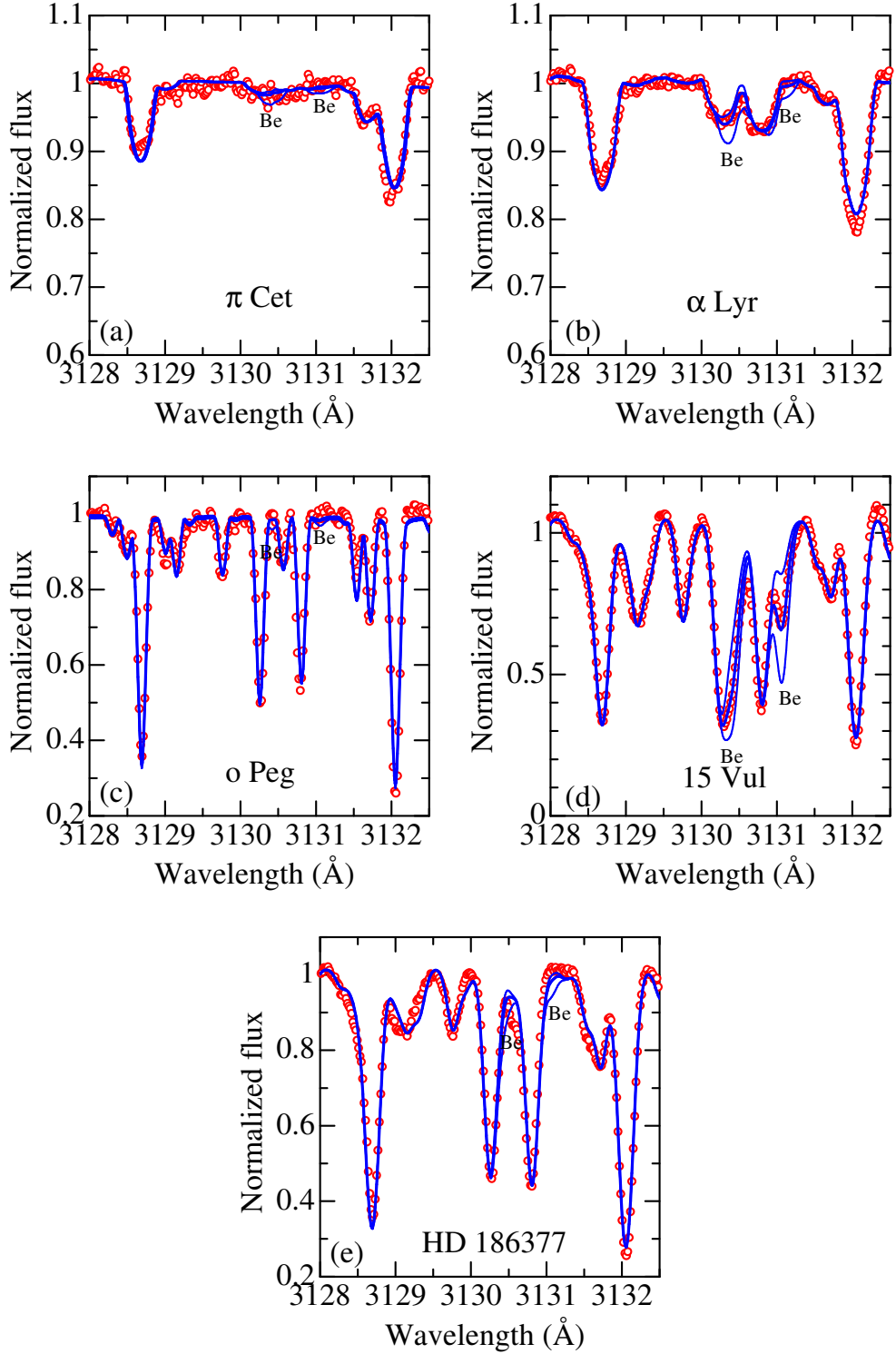}
  \end{center}
\caption{
Synthetic spectrum fitting in the 3128--3132.5~$\rm\AA$ region 
for five late~B--late~A stars. The thick solid line shows 
the best-fit theoretical spectrum corresponding to the final 
Be abundance (given in table 2), while those resulting from
perturbing $A$(Be) by $\pm 0.5$~dex around this solution are shown 
by thin solid lines. The observed spectrum is depicted plotted
by open circles. (a) $\pi$~Cet, (b) $\alpha$~Lyr, (c) $o$~Peg,
(d) 15~Vul, and (e) HD~186377.

}
\end{figure}

\newpage 

\setcounter{table}{0}
\small
\renewcommand{\arraystretch}{0.8}
\setlength{\tabcolsep}{3pt}
\begin{longtable}{ccccrccccccrrc}
\caption{Basic parameters of the program stars and the results of Be abundances.}
\hline\hline
HD\# & $T_{\rm eff}$ & $\log g$ & $v_{\rm t}$ & [Fe/H] & $v_{\rm e}\sin i$ & 
$M$ & $\log age$ & $EW_{\rm Be II\; 3131}$ & $A$(Be) & class & S/N & $ew^{\rm DL}$ & Remark \\
  & (K) & (cm~s$^{-2}$) & (km~s$^{1}$) & (dex) & (km~s$^{1}$) & 
($M_{\odot}$) & (yr) & (m$\rm\AA$) & (dex) &  &  &  (m$\rm\AA$) & \\
\hline
\endhead
\hline
\endfoot
\hline
\multicolumn{11}{l}{\hbox to 0pt{\parbox{150mm}{\footnotesize
Note. \\
Following the HD number (column 1), the fundamental stellar parameters of
the program stars, all of which were taken from Paper I, are given in 
columns 2--8: the effective temperature, logarithmic surface gravity, 
microturbulent velocity dispersion, metallicity (logarithmic Fe 
abundance relative to the Sun), projected rotational velocity, 
initial stellar mass, and logarithmic stellar age. 
Columns 9--11 present the results of our abundance analysis: 
the inversely determined equivalent width of the Be~{\sc ii}~3131.07 line, 
the logarithmic Be abundances (expressed in the usual normalization of $A$(H)=12.00), 
and the reliability class (cf. subsection 3.2), respectively, where 
the parenthesized values denote the upper limits (undetermined cases).
The spectrum quality-related quiantities (S/N ratio at $\lambda \sim 3131$~$\rm\AA$ 
and the detection-limit equivalent-width defined in subsection 3.2) are given in 
columns 12 and 13. Planet-harboring stars are denoted as ``PHS'' in column 14.
}}}
\endlastfoot
\hline
000087&  5072&  2.63& 1.35& $-$0.07&  3.6& 2.74& 8.66& 33.9 &  $-$0.22  & a &  67 & 3.1 &  \\
000360&  4850&  2.62& 1.34& $-$0.08&  1.9& 2.34& 8.86& 16.6 &  $-$0.54  & a &  63 & 3.1 &  \\
000448&  4780&  2.51& 1.32&  0.03&  2.0& 2.25& 8.99&($<$4.6)&($<-$1.15) & x &  43 & 4.6 &  \\
000587&  4893&  3.08& 1.13& $-$0.09&  1.4& 1.58& 9.36&($<$2.2)&($<-$1.24) & x &  73 & 2.2 &  \\
000645&  4880&  3.03& 1.18&  0.07&  1.8& 1.95& 9.08&  9.8 &  $-$0.52  & a &  63 & 2.7 &  \\
001239&  5114&  2.21& 1.63& $-$0.24&  4.3& 3.75& 8.28& 56.9 &  $-$0.19  & a &  70 & 3.2 &  \\
002114&  5230&  2.57& 1.57& $-$0.03&  3.3& 3.29& 8.45& 37.5 &  $-$0.24  & a &  87 & 2.4 &  \\
002952&  4844&  2.67& 1.32&  0.00&  1.9& 2.54& 8.76&($<$3.4)&($<-$1.24) & x &  56 & 3.4 & PHS \\
003421&  5287&  1.88& 2.14& $-$0.20&  6.3& 4.43& 8.13&($<$4.0)&($<-$1.81) & x &  85 & 4.0 &  \\
003546&  4882&  2.09& 1.44& $-$0.67&  3.1& 2.00& 8.95& 11.5 &  $-$1.27  & a &  83 & 2.4 &  \\
003817&  5041&  2.52& 1.40& $-$0.12&  2.5& 2.81& 8.62& 27.4 &  $-$0.42  & a &  74 & 2.6 &  \\
003856&  4766&  2.28& 1.35& $-$0.15&  1.6& 3.09& 8.55& 53.8 &  +0.02  & a &  51 & 3.9 &  \\
004188&  4844&  2.58& 1.32& $-$0.01&  2.1& 2.54& 8.75&($<$3.5)&($<-$1.27) & x &  55 & 3.5 &  \\
004398&  4892&  2.56& 1.37& $-$0.18&  1.9& 2.59& 8.72& 10.2 &  $-$0.87  & a &  60 & 3.2 &  \\
004440&  4842&  2.91& 1.15& $-$0.10&  1.5& 1.81& 9.19&($<$3.8)&($<-$1.08) & x &  43 & 3.8 &  \\
004627&  4599&  2.05& 1.40& $-$0.20&  1.8& 3.06& 8.56&($<$3.3)&($<-$1.57) & x &  59 & 3.3 &  \\
004732&  4959&  3.16& 1.12&  0.01&  1.5& 1.74& 9.24&($<$2.8)&($<-$1.08) & x &  59 & 2.8 & PHS \\
005395&  4774&  2.17& 1.40& $-$0.45&  1.2& 1.95& 9.08& 31.4 &  $-$0.57  & a &  55 & 3.5 &  \\
005608&  4854&  3.03& 1.08&  0.06&  1.4& 1.55& 9.40&($<$2.4)&($<-$1.16) & x &  74 & 2.4 & PHS \\
005722&  4893&  2.49& 1.39& $-$0.23&  1.8& 2.26& 8.95& 10.9 &  $-$0.90  & a &  77 & 2.4 &  \\
006186&  4829&  2.30& 1.35& $-$0.31&  1.8& 2.30& 8.92& 12.7 &  $-$0.94  & a &  79 & 2.5 &  \\
007087&  4908&  2.39& 1.53& $-$0.04&  2.9& 3.83& 8.28& 49.9 &  $-$0.01  & a &  77 & 2.7 &  \\
009057&  4883&  2.49& 1.37&  0.04&  2.3& 2.56& 8.78&($<$2.8)&($<-$1.42) & x &  68 & 2.8 &  \\
009408&  4746&  2.21& 1.40& $-$0.34&  1.2& 2.04& 9.00&  9.0 &  $-$1.15  & b &  58 & 3.4 &  \\
009774&  4980&  2.50& 1.60&  0.02&  5.6& 3.25& 8.46& 55.7 &  +0.13  & a &  53 & 4.8 &  \\
010348&  4931&  2.55& 1.56&  0.01&  3.1& 3.04& 8.54& 10.3 &  $-$0.82  & a &  68 & 3.0 &  \\
010761&  4952&  2.43& 1.43& $-$0.05&  2.6& 3.04& 8.53& 11.3 &  $-$0.87  & a &  65 & 3.1 &  \\
010975&  4866&  2.47& 1.37& $-$0.17&  1.7& 2.19& 8.94&  4.2 &  $-$1.33  & b &  71 & 2.6 &  \\
011037&  4862&  2.45& 1.33& $-$0.14&  1.7& 2.30& 8.88& 17.9 &  $-$0.63  & a &  65 & 2.9 &  \\
011949&  4845&  2.85& 1.17& $-$0.10&  1.3& 2.17& 8.94& 24.8 &  $-$0.19  & a &  65 & 2.7 &  \\
012139&  4833&  2.53& 1.36& $-$0.09&  1.7& 2.45& 8.87&($<$2.8)&($<-$1.43) & x &  68 & 2.8 &  \\
012339&  5011&  2.52& 1.51& $-$0.03&  2.8& 3.19& 8.48&($<$2.5)&($<-$1.53) & x &  80 & 2.5 &  \\
013468&  4893&  2.54& 1.34& $-$0.16&  1.6& 2.31& 8.92& 24.2 &  $-$0.43  & a &  61 & 3.2 &  \\
013994&  4974&  2.44& 1.83& $-$0.11& 10.1& 3.84& 8.28& 49.6 &  $-$0.06  & a &  49 & 7.2 &  \\
014129&  4936&  2.61& 1.37& $-$0.01&  2.5& 2.70& 8.68&($<$2.8)&($<-$1.39) & x &  64 & 2.8 &  \\
014770&  4977&  2.47& 1.47&  0.01&  2.5& 3.03& 8.54& 29.7 &  $-$0.33  & a & 121 & 1.6 &  \\
015779&  4846&  2.63& 1.26&  0.00&  1.8& 2.49& 8.78&($<$3.2)&($<-$1.29) & x &  62 & 3.2 & PHS \\
015920&  5061&  2.74& 1.33& $-$0.06&  3.2& 2.63& 8.71& 33.7 &  $-$0.15  & a &  53 & 3.8 &  \\
016400&  4785&  2.35& 1.33& $-$0.06&  1.8& 2.43& 8.82&($<$3.6)&($<-$1.39) & x &  55 & 3.6 & PHS \\
016901&  5624&  1.42& 3.17&  0.00&  8.7& 4.03& 8.21&($<$6.1)&($<-$1.72) & x &  74 & 6.1 &  \\
017656&  5100&  2.67& 1.37& $-$0.06&  2.6& 2.73& 8.66& 12.7 &  $-$0.74  & a &  81 & 2.5 &  \\
017824&  5051&  2.82& 1.19& $-$0.04&  3.2& 2.37& 8.83& 41.4 &  +0.05  & a &  55 & 3.5 &  \\
018474&  5013&  2.38& 1.42& $-$0.23&  2.6& 3.59& 8.33&($<$3.2)&($<-$1.58) & x &  61 & 3.2 &  \\
018953&  5029&  2.93& 1.23&  0.14&  2.5& 2.53& 8.74&($<$3.2)&($<-$1.12) & x &  60 & 3.2 &  \\
018970&  4791&  2.44& 1.30& $-$0.07&  2.1& 2.44& 8.81&  1.6&  $-$1.70 & c &  62 & 3.2 &  \\
019476&  4933&  2.82& 1.24&  0.14&  2.3& 2.36& 8.83&($<$2.9)&($<-$1.19) & x &  65 & 2.9 &  \\
092125&  5468&  2.22& 2.07&  0.03&  6.4& 3.72& 8.30&($<$6.6)&($<-$1.34) & x &  52 & 6.6 &  \\
093291&  5039&  2.74& 1.28& $-$0.10&  2.8& 2.43& 8.79&($<$4.7)&($<-$1.16) & x &  39 & 4.7 &  \\
106057&  4956&  2.64& 1.35& $-$0.10&  1.9& 3.06& 8.52&($<$3.1)&($<-$1.37) & x &  57 & 3.1 &  \\
106714&  4933&  2.57& 1.37& $-$0.18&  1.8& 2.50& 8.79& 16.5 &  $-$0.65  & a &  60 & 3.1 &  \\
107383&  4841&  2.51& 1.38& $-$0.28&  1.8& 3.14& 8.49&($<$3.0)&($<-$1.48) & x &  56 & 3.0 & PHS \\
107950&  5171&  2.60& 1.63&  0.01&  5.4& 3.36& 8.42&($<$4.6)&($<-$1.25) & x &  51 & 4.6 &  \\
108225&  4969&  2.71& 1.27&  0.04&  2.8& 2.50& 8.77&  7.0 &  $-$0.91  & b &  64 & 3.1 &  \\
109272&  5104&  3.22& 1.13& $-$0.26&  1.6& 1.79& 9.16&  6.5 &  $-$0.80  & b &  52 & 3.0 &  \\
109317&  4866&  2.61& 1.38& $-$0.05&  1.7& 2.41& 8.82&($<$3.6)&($<-$1.27) & x &  52 & 3.6 &  \\
110646&  5067&  3.05& 1.21& $-$0.45&  1.1& 1.81& 9.12&($<$2.2)&($<-$1.45) & x &  70 & 2.2 &  \\
111028&  4881&  3.27& 1.03& $-$0.05&  0.7& 1.41& 9.53&($<$2.6)&($<-$1.03) & x &  57 & 2.6 &  \\
113095&  4961&  2.68& 1.37& $-$0.07&  2.1& 2.59& 8.73&  6.6 &  $-$1.00  & b &  55 & 3.1 &  \\
113226&  5044&  2.63& 1.41&  0.07&  3.0& 2.70& 8.68& 16.6 &  $-$0.55  & a &  45 & 4.4 &  \\
114256&  4858&  2.68& 1.34&  0.04&  2.0& 2.51& 8.77& 12.3 &  $-$0.62  & b &  42 & 4.2 &  \\
115659&  5019&  2.47& 1.47& $-$0.06&  3.9& 2.94& 8.57& 54.0 &  +0.04  & a &  44 & 4.5 &  \\
116957&  4898&  2.63& 1.33& $-$0.10&  1.5& 2.39& 8.87&  8.7 &  $-$0.88  & b &  53 & 3.3 &  \\
117818&  4811&  2.31& 1.34& $-$0.34&  1.9& 2.05& 9.06& 18.2 &  $-$0.76  & a &  51 & 3.5 &  \\
118219&  4831&  2.34& 1.33& $-$0.34&  1.5& 2.51& 8.74&  9.1 &  $-$1.10  & b &  51 & 3.7 &  \\
119126&  4796&  2.33& 1.34& $-$0.12&  1.6& 2.38& 8.85&  2.8 &  $-$1.54  & b &  73 & 2.6 &  \\
119605&  5456&  1.96& 1.95& $-$0.31&  6.6& 4.04& 8.19& 10.2 &  $-$1.38  & b &  57 & 6.1 &  \\
120048&  5014&  2.79& 1.22&  0.11&  3.5& 2.71& 8.66& 54.5 &  +0.32  & a &  58 & 3.6 &  \\
120420&  4791&  2.63& 1.26& $-$0.20&  0.7& 2.25& 8.90& 10.6 &  $-$0.78  & a &  80 & 2.2 &  \\
120787&  4843&  2.31& 1.34& $-$0.38&  1.8& 2.02& 9.08& 22.5 &  $-$0.67  & a &  49 & 4.1 &  \\
125454&  4848&  2.56& 1.39& $-$0.10&  1.8& 2.47& 8.82&  6.2 &  $-$1.06  & b &  56 & 3.5 &  \\
126218&  5025&  2.50& 1.58&  0.12&  3.5& 3.15& 8.48& 26.0 &  $-$0.35  & a &  43 & 5.1 &  \\
127243&  4893&  2.21& 1.48& $-$0.77&  3.4& 1.92& 9.08& 11.3 &  $-$1.25  & a &  77 & 2.6 &  \\
129312&  4993&  2.53& 1.62&  0.01&  5.9& 4.20& 8.15& 61.3 &  +0.22  & a &  71 & 3.9 &  \\
129336&  4901&  2.54& 1.33& $-$0.25&  2.3& 2.68& 8.67&($<$2.2)&($<-$1.62) & x &  85 & 2.2 &  \\
129944&  4892&  2.50& 1.32& $-$0.26&  2.0& 2.59& 8.70&($<$2.5)&($<-$1.58) & x &  69 & 2.5 &  \\
129972&  4976&  2.69& 1.43& $-$0.01&  3.2& 2.68& 8.69& 23.2 &  $-$0.35  & a &  77 & 2.5 &  \\
130952&  4750&  2.34& 1.35& $-$0.40&  4.3& 1.85& 9.10& 10.7 &  $-$1.01  & a &  81 & 2.7 &  \\
131530&  4962&  2.72& 1.33&  0.00&  2.7& 2.72& 8.67& 15.9 &  $-$0.52  & a &  51 & 3.9 &  \\
132146&  5012&  2.29& 1.60& $-$0.06&  2.8& 3.45& 8.39&  5.1 &  $-$1.34  & b &  72 & 2.8 &  \\
133208&  5001&  2.35& 1.61& $-$0.07&  3.1& 3.42& 8.40& 50.9 &  $-$0.08  & a & 140 & 1.5 &  \\
133392&  4903&  2.69& 1.32&  0.09&  2.2& 2.49& 8.77&($<$2.3)&($<-$1.38) & x &  82 & 2.3 &  \\
134190&  4841&  2.28& 1.40& $-$0.41&  1.5& 2.03& 8.99&($<$2.0)&($<-$1.85) & x &  92 & 2.0 &  \\
136512&  4749&  2.34& 1.39& $-$0.29&  2.3& 2.13& 8.93&  9.2 &  $-$1.04  & b &  59 & 3.3 & PHS \\
136956&  5031&  2.61& 1.54&  0.08&  3.0& 3.78& 8.27&($<$4.5)&($<-$1.18) & x &  48 & 4.5 &  \\
138716&  4830&  3.14& 1.05&  0.00&  1.3& 1.44& 9.51&($<$2.4)&($<-$1.10) & x &  66 & 2.4 &  \\
138852&  4900&  2.55& 1.36& $-$0.22&  2.0& 2.21& 8.98& 17.2 &  $-$0.64  & a &  46 & 3.9 &  \\
138905&  4822&  2.56& 1.27& $-$0.30&  1.5& 2.15& 9.01& 46.2 &  $-$0.03  & a &  62 & 2.9 &  \\
139641&  4907&  2.75& 1.16& $-$0.53&  0.7& 1.43& 9.50&  5.2 &  $-$1.22  & b &  68 & 2.3 &  \\
141680&  4770&  2.32& 1.34& $-$0.24&  1.9& 2.17& 8.95&  2.8 &  $-$1.58  & b &  88 & 2.1 & PHS \\
142091&  4877&  3.21& 1.04&  0.10&  1.2& 1.51& 9.43&($<$1.6)&($<-$1.22) & x & 107 & 1.6 & PHS \\
142198&  4760&  2.35& 1.39& $-$0.27&  1.7& 2.13& 9.02&($<$2.4)&($<-$1.64) & x &  79 & 2.4 &  \\
142531&  4961&  2.78& 1.37&  0.05&  2.4& 2.64& 8.70&($<$3.8)&($<-$1.14) & x &  50 & 3.8 &  \\
143553&  4805&  2.85& 1.17& $-$0.23&  0.7& 1.75& 9.20& 28.7 &  $-$0.14  & a &  54 & 3.0 &  \\
144608&  5266&  2.54& 1.60& $-$0.09&  3.5& 3.27& 8.45& 26.8 &  $-$0.49  & a &  89 & 2.3 &  \\
145001&  5119&  2.90& 1.57&  0.04&  8.8& 3.17& 8.49& 38.5 &  +0.04  & a & 112 & 2.9 &  \\
146791&  4931&  2.69& 1.34& $-$0.07&  2.6& 2.52& 8.78& 18.2 &  $-$0.48  & a &  78 & 2.5 &  \\
147677&  4978&  2.90& 1.28&  0.10&  2.4& 2.36& 8.83&($<$1.6)&($<-$1.45) & x & 118 & 1.6 &  \\
147700&  4843&  2.48& 1.31& $-$0.11&  1.7& 2.35& 8.89& 29.1 &  $-$0.32  & a &  70 & 2.7 &  \\
148387&  5055&  2.82& 1.34& $-$0.04&  2.6& 2.55& 8.74&  6.2 &  $-$0.97  & b &  59 & 3.2 &  \\
148604&  5120&  2.90& 0.98& $-$0.16&  2.4& 2.48& 8.76& 35.8 &  $-$0.07  & a &  62 & 3.6 &  \\
148786&  5110&  2.69& 1.52&  0.17&  5.2& 2.96& 8.55& 26.1 &  $-$0.25  & a &  80 & 2.8 &  \\
150030&  4850&  2.10& 1.81& $-$0.09&  4.0& 4.02& 8.22&  4.8 &  $-$1.44  & b &  69 & 3.3 &  \\
150997&  5045&  2.79& 1.26& $-$0.15&  2.7& 2.41& 8.80&($<$1.4)&($<-$1.68) & x & 135 & 1.4 &  \\
152815&  4859&  2.43& 1.35& $-$0.21&  1.8& 2.19& 8.93& 11.1 &  $-$0.91  & a &  63 & 3.0 &  \\
154084&  4862&  2.62& 1.41& $-$0.16&  2.0& 2.39& 8.89&($<$2.7)&($<-$1.43) & x &  68 & 2.7 &  \\
154779&  5064&  2.75& 1.44&  0.12&  3.1& 2.79& 8.63&  4.6 &  $-$1.08  & b &  51 & 3.8 &  \\
156874&  4982&  2.85& 1.32&  0.00&  2.5& 2.53& 8.76&($<$2.3)&($<-$1.35) & x &  80 & 2.3 &  \\
156891&  4981&  2.95& 1.30&  0.13&  3.0& 2.44& 8.79&($<$3.2)&($<-$1.10) & x &  62 & 3.2 &  \\
157527&  5090&  2.96& 1.30&  0.07&  3.2& 2.49& 8.77&($<$3.8)&($<-$1.07) & x &  55 & 3.8 &  \\
158974&  4901&  2.32& 1.43& $-$0.07&  3.0& 2.74& 8.66& 20.3 &  $-$0.62  & a &  86 & 2.3 &  \\
159181&  5153&  1.50& 2.69& $-$0.15&  7.5& 4.65& 8.09&($<$6.7)&($<-$1.73) & x &  68 & 6.7 &  \\
159353&  4919&  2.76& 1.32&  0.00&  2.2& 2.69& 8.69&  6.8 &  $-$0.89  & b &  65 & 2.8 &  \\
160781&  4593&  2.10& 1.62& $-$0.02&  4.9& 4.99& 8.00&115.0 &  +1.18  & a &  40 & 6.5 &  \\
161178&  4766&  2.33& 1.32& $-$0.20&  1.8& 2.14& 8.94&($<$4.7)&($<-$1.33) & x &  40 & 4.7 &  \\
162076&  5018&  2.98& 1.24&  0.04&  3.1& 2.27& 8.89& 40.7 &  +0.17  & a &  63 & 3.3 &  \\
163532&  4689&  2.17& 1.44& $-$0.06&  2.2& 3.17& 8.55&($<$4.4)&($<-$1.36) & x &  44 & 4.4 &  \\
163917&  4928&  2.63& 1.46&  0.13&  3.2& 3.04& 8.52&($<$3.4)&($<-$1.23) & x &  60 & 3.4 & PHS \\
165760&  4962&  2.52& 1.41& $-$0.01&  2.8& 2.82& 8.63& 11.0 &  $-$0.82  & a &  75 & 2.6 &  \\
167042&  4943&  3.28& 1.07&  0.00&  0.7& 1.50& 9.45&($<$2.5)&($<-$1.05) & x &  66 & 2.5 & PHS \\
167768&  4895&  2.13& 1.44& $-$0.70&  4.4& 2.07& 8.90& 24.0 &  $-$0.89  & a &  76 & 2.8 &  \\
168656&  5045&  2.66& 1.30& $-$0.06&  2.6& 2.86& 8.60& 25.3 &  $-$0.36  & a &  85 & 2.3 &  \\
168723&  4972&  3.12& 1.17& $-$0.18&  1.5& 1.84& 9.14&($<$2.1)&($<-$1.30) & x &  82 & 2.1 &  \\
170474&  4978&  2.83& 1.29&  0.02&  2.1& 2.47& 8.78&($<$2.4)&($<-$1.33) & x &  75 & 2.4 &  \\
171391&  5057&  2.79& 1.23& $-$0.02&  3.2& 2.84& 8.62& 43.1 &  +0.06  & a &  69 & 2.9 &  \\
174980&  5008&  2.71& 1.41&  0.10&  2.8& 2.81& 8.62&($<$4.3)&($<-$1.12) & x &  45 & 4.3 &  \\
176598&  5018&  2.83& 1.21&  0.03&  4.3& 2.52& 8.76& 57.1 &  +0.35  & a &  56 & 4.1 &  \\
176707&  4777&  2.27& 1.38& $-$0.29&  1.0& 2.01& 9.03& 19.4 &  $-$0.72  & a &  70 & 2.6 &  \\
177241&  4906&  2.70& 1.36&  0.01&  2.1& 2.63& 8.71&($<$3.4)&($<-$1.23) & x &  57 & 3.4 &  \\
177249&  5251&  2.55& 1.65&  0.00&  5.7& 3.12& 8.51& 22.2 &  $-$0.55  & a & 106 & 2.3 &  \\
180540&  4951&  2.34& 1.76& $-$0.08&  5.2& 4.34& 8.11& 25.0 &  $-$0.53  & a &  55 & 4.6 &  \\
180711&  4885&  2.62& 1.38& $-$0.13&  1.6& 2.32& 8.91& 14.2 &  $-$0.65  & a &  67 & 2.7 &  \\
181276&  4986&  2.78& 1.32&  0.04&  2.0& 2.41& 8.81&  8.8 &  $-$0.77  & a &  81 & 2.3 &  \\
182694&  5067&  2.63& 1.37& $-$0.04&  3.3& 2.67& 8.69& 40.0 &  $-$0.10  & a &  83 & 2.5 &  \\
182762&  4872&  2.57& 1.34& $-$0.07&  1.8& 2.42& 8.82&  9.4 &  $-$0.85  & a &  69 & 2.7 &  \\
183491&  4901&  2.63& 1.40&  0.11&  2.7& 3.07& 8.51&  1.9&  $-$1.49 & c &  67 & 3.2 &  \\
184010&  5011&  3.17& 1.16& $-$0.14&  1.3& 1.82& 9.16&  9.2 &  $-$0.60  & a &  81 & 1.9 &  \\
185018&  5467&  1.85& 2.31& $-$0.10&  7.0& 4.76& 8.06&($<$5.2)&($<-$1.66) & x &  73 & 5.2 &  \\
185194&  4978&  2.44& 1.54&  0.03&  3.2& 3.09& 8.52& 52.6 &  +0.05  & a &  70 & 3.0 &  \\
185351&  5006&  3.16& 1.15&  0.00&  1.3& 1.76& 9.23&($<$1.9)&($<-$1.27) & x &  90 & 1.9 &  \\
185467&  4937&  2.70& 1.45&  0.13&  2.6& 2.83& 8.61&($<$4.8)&($<-$1.04) & x &  39 & 4.8 &  \\
185758&  5535&  2.39& 1.87&  0.01&  5.4& 4.11& 8.18&  3.5 &  $-$1.55  & b &  92 & 3.2 &  \\
185958&  4876&  2.22& 2.08&  0.02&  8.6& 4.33& 8.11&  4.4&  $-$1.38 & c &  60 & 5.1 &  \\
186675&  4953&  2.46& 1.47& $-$0.08&  2.9& 2.74& 8.66&($<$1.7)&($<-$1.74) & x & 116 & 1.7 &  \\
187739&  4771&  2.71& 1.03& $-$0.19&  2.3& 2.01& 9.08& 67.7 &  +0.52  & a &  54 & 4.2 &  \\
188310&  4802&  2.72& 1.42& $-$0.18&  3.8& 2.29& 8.89&($<$3.6)&($<-$1.23) & x &  56 & 3.6 & PHS \\
188650&  5450&  1.79& 2.17& $-$0.67&  7.2& 4.64& 8.07&($<$3.4)&($<-$2.08) & x &  99 & 3.4 &  \\
188947&  4866&  2.69& 1.35&  0.07&  2.1& 2.56& 8.74& 11.2 &  $-$0.65  & a &  60 & 3.2 &  \\
189127&  4760&  2.28& 1.41& $-$0.22&  1.8& 2.31& 8.92& 19.5 &  $-$0.68  & a &  54 & 3.7 &  \\
192787&  5025&  2.86& 1.25& $-$0.07&  2.1& 2.47& 8.77& 16.7 &  $-$0.47  & a &  90 & 2.0 &  \\
192879&  4886&  2.62& 1.37& $-$0.09&  1.6& 2.47& 8.83&  3.0&  $-$1.36 & c &  44 & 4.4 &  \\
192944&  4981&  2.48& 1.48& $-$0.06&  3.7& 3.41& 8.40& 37.6 &  $-$0.20  & a &  61 & 3.7 &  \\
192947&  5046&  2.90& 1.32&  0.03&  3.0& 2.43& 8.80& 31.7 &  $-$0.05  & a &  65 & 3.1 &  \\
194013&  4906&  2.63& 1.32& $-$0.07&  2.3& 2.36& 8.84&  5.9 &  $-$1.05  & b &  69 & 2.6 &  \\
194577&  5028&  2.68& 1.34& $-$0.02&  4.6& 3.35& 8.43&($<$3.1)&($<-$1.35) & x &  68 & 3.1 &  \\
196857&  4878&  2.55& 1.44& $-$0.27&  1.7& 2.15& 9.01& 23.4 &  $-$0.49  & a &  52 & 3.7 &  \\
199665&  4985&  2.84& 1.19& $-$0.05&  2.7& 2.25& 8.90&($<$2.7)&($<-$1.31) & x &  70 & 2.7 & PHS \\
200039&  4965&  2.67& 1.36& $-$0.13&  2.0& 2.62& 8.70&  4.2 &  $-$1.24  & b &  49 & 3.7 &  \\
201381&  4951&  2.77& 1.30& $-$0.04&  2.4& 2.35& 8.85& 13.0 &  $-$0.60  & a &  61 & 3.1 &  \\
203222&  5067&  2.78& 1.29& $-$0.02&  2.2& 2.49& 8.77&  5.7 &  $-$1.02  & b &  73 & 2.5 &  \\
203387&  5244&  3.07& 1.26&  0.07&  5.3& 2.79& 8.63& 26.5 &  $-$0.12  & a &  63 & 3.8 &  \\
204381&  5100&  2.84& 1.33& $-$0.06&  2.4& 2.47& 8.78& 14.0 &  $-$0.59  & a &  60 & 3.2 &  \\
204771&  4967&  2.93& 1.26&  0.09&  2.2& 2.44& 8.79&($<$3.4)&($<-$1.09) & x &  53 & 3.4 &  \\
205072&  4995&  2.72& 1.34& $-$0.14&  2.1& 2.41& 8.80&($<$4.5)&($<-$1.19) & x &  30 & 4.5 &  \\
205435&  5114&  3.00& 1.20& $-$0.10&  3.1& 2.33& 8.85& 31.8 &  $-$0.07  & a &  59 & 3.4 &  \\
206356&  4938&  2.80& 1.28&  0.11&  2.3& 2.55& 8.74& 15.7 &  $-$0.42  & a &  52 & 3.8 &  \\
206453&  5038&  2.43& 1.48& $-$0.38&  2.4& 2.97& 8.53& 13.4 &  $-$0.95  & a &  66 & 2.9 &  \\
209396&  4999&  2.81& 1.30&  0.04&  3.1& 2.46& 8.79& 12.7 &  $-$0.58  & a &  71 & 2.8 &  \\
210354&  4793&  2.36& 1.39& $-$0.22&  3.0& 1.92& 9.12&($<$3.1)&($<-$1.52) & x &  62 & 3.1 &  \\
210434&  4949&  2.93& 1.36&  0.12&  2.7& 2.29& 8.87& 16.3 &  $-$0.33  & a &  70 & 2.6 &  \\
210702&  4967&  3.19& 1.10&  0.01&  2.0& 1.68& 9.28&($<$2.2)&($<-$1.17) & x &  76 & 2.2 & PHS \\
210807&  5071&  2.58& 1.57& $-$0.10&  6.7& 3.50& 8.37& 31.6 &  $-$0.31  & a &  51 & 5.2 &  \\
211391&  4909&  2.57& 1.36&  0.09&  2.3& 2.78& 8.64& 11.8 &  $-$0.69  & a &  59 & 3.4 &  \\
211434&  5082&  2.70& 1.37& $-$0.26&  1.9& 2.53& 8.73& 13.5 &  $-$0.76  & a &  77 & 2.3 &  \\
211554&  5043&  2.41& 1.63&  0.05&  3.8& 4.26& 8.12& 26.8 &  $-$0.43  & a &  70 & 3.1 &  \\
212271&  5002&  2.90& 1.21&  0.10&  2.6& 2.50& 8.76& 44.4 &  +0.22  & a &  50 & 3.9 &  \\
212320&  5075&  2.59& 1.46& $-$0.11&  3.5& 2.84& 8.61& 54.6 &  +0.08  & a &  76 & 2.8 &  \\
212430&  4954&  2.56& 1.39& $-$0.17&  2.0& 3.17& 8.49& 16.7 &  $-$0.65  & a &  58 & 3.2 &  \\
212496&  4710&  2.43& 1.22& $-$0.33&  1.4& 1.85& 9.12&($<$3.2)&($<-$1.47) & x &  57 & 3.2 &  \\
213789&  5010&  2.73& 1.37& $-$0.06&  2.6& 2.77& 8.64&  8.8 &  $-$0.85  & a &  67 & 2.9 &  \\
213930&  5011&  2.87& 1.34&  0.12&  4.2& 2.75& 8.65& 58.9 &  +0.45  & a &  54 & 4.3 &  \\
213986&  4928&  2.83& 1.27&  0.08&  2.1& 2.50& 8.76&  7.6 &  $-$0.77  & b &  54 & 3.7 &  \\
214567&  4989&  2.69& 1.33& $-$0.21&  2.6& 2.57& 8.71& 10.4 &  $-$0.85  & a &  83 & 2.3 &  \\
214878&  5041&  2.85& 1.29&  0.04&  2.4& 2.62& 8.71&  1.7&  $-$1.48 & c &  58 & 3.3 &  \\
215030&  4731&  2.41& 1.25& $-$0.49&  0.7& 1.83& 9.24&  4.8 &  $-$1.37  & b &  63 & 2.8 &  \\
215373&  5007&  2.69& 1.39&  0.10&  3.7& 2.66& 8.69& 12.0 &  $-$0.65  & a &  66 & 3.1 &  \\
215721&  4829&  2.23& 1.39& $-$0.48&  1.0& 1.95& 9.07&  8.7 &  $-$1.23  & b &  65 & 2.9 &  \\
215943&  4878&  2.68& 1.33& $-$0.04&  1.9& 2.45& 8.84&($<$3.2)&($<-$1.28) & x &  60 & 3.2 &  \\
216131&  5000&  2.69& 1.24& $-$0.05&  2.2& 2.49& 8.77&  9.7 &  $-$0.82  & a &  64 & 2.9 &  \\
217264&  4946&  2.80& 1.27&  0.12&  2.0& 2.55& 8.74&($<$2.5)&($<-$1.28) & x &  75 & 2.5 &  \\
217703&  4890&  2.91& 1.16& $-$0.17&  0.9& 1.98& 9.05& 20.0 &  $-$0.32  & a &  57 & 3.0 &  \\
218527&  4935&  2.57& 1.33& $-$0.34&  3.7& 2.11& 9.03&($<$2.9)&($<-$1.53) & x &  89 & 2.9 &  \\
219139&  4860&  2.50& 1.38& $-$0.19&  2.0& 2.29& 8.88&  4.6 &  $-$1.27  & b &  64 & 2.9 &  \\
219615&  4802&  2.25& 1.37& $-$0.62&  1.7& 1.67& 9.20& 24.8 &  $-$0.73  & a &  63 & 3.0 &  \\
219945&  4874&  2.61& 1.36& $-$0.10&  1.7& 2.57& 8.77&  5.2 &  $-$1.12  & b &  70 & 2.7 &  \\
221345&  4813&  2.63& 1.43& $-$0.24&  2.6& 2.20& 8.93&($<$3.1)&($<-$1.38) & x &  65 & 3.1 & PHS \\
222093&  4853&  2.56& 1.38& $-$0.12&  2.0& 2.28& 8.89&($<$3.8)&($<-$1.29) & x &  51 & 3.8 &  \\
222387&  5055&  2.81& 1.22& $-$0.11&  2.7& 2.79& 8.63& 38.8 &  $-$0.03  & a &  44 & 5.1 &  \\
222574&  5523&  1.99& 2.20&  0.04&  7.0& 4.23& 8.13&($<$5.4)&($<-$1.54) & x &  69 & 5.4 &  \\
223252&  5031&  2.72& 1.34& $-$0.03&  2.8& 2.52& 8.76& 13.0 &  $-$0.66  & a &  65 & 2.9 &  \\
224533&  5030&  2.73& 1.28& $-$0.01&  2.6& 2.54& 8.75& 13.2 &  $-$0.64  & a &  73 & 2.6 &  \\
\end{longtable}

\newpage 

\setcounter{table}{1}
\setlength{\tabcolsep}{3pt}
\begin{table}[h]
\small
\caption{Error simulation based on artificial spectra.} 
\begin{center}
\begin{tabular}
{r rcr rcr rcr}\hline \hline
$A_{\rm given}$ & 
$\langle A \rangle$  & $\sigma$ & $N$ &
$\langle A \rangle$  & $\sigma$ & $N$ &
$\langle A \rangle$  & $\sigma$ & $N$ \\
    & \multicolumn{3}{c}{(S/N = 100)} & \multicolumn{3}{c}{(S/N = 50)} & \multicolumn{3}{c}{(S/N = 20)} \\
\hline
 +1.0 &  +0.999 & 0.010 &  100  &   +0.999 & 0.020 &  100  &   +0.996 & 0.050 &  100\\
 +0.5 &  +0.502 & 0.012 &  100  &   +0.504 & 0.024 &  100  &   +0.510 & 0.060 &  100\\
 0.0 &  0.000 & 0.014 &  100  &   +0.001 & 0.028 &  100  &   0.000 & 0.071 &  100\\
$-$0.5 & $-$0.506 & 0.035 &  100  &  $-$0.515 & 0.072 &  100  &  $-$0.551 & 0.195 &   99\\
$-$1.0 & $-$1.012 & 0.079 &  100  &  $-$1.024 & 0.145 &   98  &  $-$1.004 & 0.271 &   86\\
$-$1.5 & $-$1.500 & 0.236 &   92  &  $-$1.402 & 0.451 &   72  &  $-$1.107 & 0.358 &   61\\
\hline
\end{tabular}
\end{center}
Note. \\
Experiment of Be abundance derivations based on many artificial spectra computed 
with different input Be abundances (+1.0, +0.5, 0.0, $-0.5$, $-1.0$, and $-1.5$) and
different S/N ratios (100, 50, and 20). Shown here are the average of resulting 
abundances ($\langle A \rangle$), the standard deviation ($\sigma$; in dex), 
and the number of successful determinations ($N$; among 100 trials) for each case.  
See appendix 1 for more details. 
\end{table}

\setcounter{table}{2}
\setlength{\tabcolsep}{3pt}
\begin{table}[h]
\small
\caption{Be abundance results of five late~B--late~A stars.} 
\begin{center}
\begin{tabular}
{cccccccl}\hline \hline
Star & HD\# & Sp.Type & $T_{\rm eff}$ & $\log g$ & $v_{\rm t}$ & $A$(Be) & Remark \\
\hline
$\pi$~Cet    &  17081 & B7~V   & 13063 & 3.72 & 1.0 &  +0.78:  & near to upper limit \\
$\alpha$~Lyr & 172167 & A0~V   &  9435 & 3.99 & 2.0 &  +0.85:  & near to upper limit \\
$o$~Peg      & 214994 & A1~IV  &  9453 & 3.64 & 2.0 &  +0.05:  & near to upper limit \\
15~Vul       & 189849 & A4~III &  7870 & 3.62 & 4.0 &  +1.47   & clearly detected \\
HD~186377    & 186377 & A5~III &  7733 & 2.40 & 4.0 & $-0.38:$ & near to upper limit, late-A giant\\
\hline
\end{tabular}
\end{center}
Note. \\
See appendix 2 for details. The ``:'' (colon) attached to $A$(Be) denotes ``uncertain value.'' 
\end{table}

\end{document}